\documentclass[12pt]{article}

\usepackage{graphicx}
\usepackage{natbib} 
\usepackage{url} 

\newcommand{\blind}{0}

\usepackage[utf8]{inputenc}
\usepackage[margin=1in]{geometry}
\usepackage{amssymb}
\usepackage{amsmath}
\usepackage{bm}
\usepackage{setspace}
\usepackage{url}
\usepackage{caption}
\usepackage{subcaption}
\usepackage{eso-pic}
\usepackage{float}
\usepackage{enumitem}
\usepackage{multirow}
\usepackage{xr-hyper}
\usepackage{hyperref}
\usepackage{tablefootnote}
\usepackage{subfiles}
\usepackage{tikz}
\usetikzlibrary{arrows,shapes, automata}
\usepackage{parskip}

\usepackage{algorithm}
\usepackage{algpseudocode}
\usepackage[section]{placeins}

\makeatletter

\def\spacingset#1{\renewcommand{\baselinestretch}%
{#1}\small\normalsize} \spacingset{1}

\begin{document}


\if0\blind
{
  \title{\bf Automated threshold selection and associated inference uncertainty for univariate extremes}
  \author{Conor Murphy\thanks{This paper is based on work completed while Conor Murphy was part of the EPSRC funded STOR-i centre for doctoral training (EP/S022252/1), with part-funding from Shell Research Ltd.}, Jonathan A. Tawn\hspace{.2cm}\\
    Department of Mathematics and Statistics, Lancaster University\\
    and \\
     Zak Varty\\
    Department of Mathematics, Imperial College London}
  \maketitle
} \fi

\if1\blind
{
  \bigskip
  \bigskip
  \bigskip
  \begin{center}
    {\LARGE\bf Automated threshold selection and associated inference uncertainty for univariate extremes}
\end{center}
  \medskip
} \fi

\bigskip
\begin{abstract}
Threshold selection is a fundamental problem in any threshold-based extreme value analysis. While models are asymptotically motivated, selecting an appropriate threshold for finite samples is difficult and highly subjective through standard methods. Inference for high quantiles can also be highly sensitive to the choice of threshold. Too low a threshold choice leads to bias in the fit of the extreme value model, while too high a choice leads to unnecessary additional uncertainty in the estimation of model parameters. We develop a novel methodology for automated threshold selection that directly tackles this bias-variance trade-off. We also develop a method to account for the uncertainty in the threshold estimation and propagate this uncertainty through to high quantile inference. Through a simulation study, we demonstrate the effectiveness of our method for threshold selection and subsequent extreme quantile estimation, relative to the leading existing methods, and show how the method's effectiveness is not sensitive to the tuning parameters. We apply our method to the well-known, troublesome example of the River Nidd dataset.
\end{abstract}

\noindent%
\vfill

\newpage
\spacingset{2} 
\section{Introduction}\label{section: intro}
An inherent challenge in risk modelling is the estimation of high quantiles, known as \textit{return levels}, beyond observed values. Such inference is important for designing policies or protections against future extreme events, e.g., in finance or hydrology \citep{smith2003statistics, coles2003fully}. Extreme value methods achieve this extrapolation by using asymptotically exact models to approximate the tail of a distribution above a high, within-sample, threshold $u$. The choice of this threshold is fundamental in providing meaningful inference.
Here, we develop novel methods for automatic selection of the threshold and for propagating the uncertainty in this selection into return level inferences.

Throughout, we assume that all data are realisations  of an independent and identically-distributed (iid) univariate continuous random variable $X$ with unknown distribution function $F$, with upper endpoint $x^F := \text{sup}\{x : F(x) < 1\}$. Under weak conditions, \citet{pickands1975statistical} shows that for $X>u$, with $u < x^F$, 
the distribution of the rescaled excess $Y = X - u$, converges to the generalised Pareto distribution (GPD) as $u \rightarrow x^F$. 
To use this limit result in practice, 
a within-sample threshold $u$ is chosen, above which this limit result is treated as exact. Specifically, whatever the form of $F$, the excesses $Y$ of $u$ are modelled by the single flexible GPD$(\sigma_u, \xi)$ family, with distribution function
\begin{equation}
    H(y; \sigma_u, \xi) = 
    1 - \left(1 + \xi y/\sigma_u\right)_+^{-1/\xi}, 
    \label{eqn: gpd}
\end{equation}
with $y > 0$, $w_+ = \text{max}(w,0)$, $(\sigma_u,\xi) \in \mathbb{R}_+\times\mathbb{R}$ being scale and shape parameters. The exponential distribution
arises when $\xi = 0$, 
i.e., as $\xi \rightarrow 0$ in distribution~\eqref{eqn: gpd}, whereas for $\xi>0$, the distribution tail decay is polynomial.
For $\xi < 0$, $X$ has a finite upper end-point at $u - \sigma_u/\xi$ but is unbounded above for $\xi \geq 0$. 
To estimate the $(1-p)$\textsuperscript{th} quantile, $x_p$, of $X$, for 
$p<\lambda_u := \mathbb{P}(X > u)$, we can solve $\hat{F}(x_p) = 1-p$, where  
$\hat{F}(x_p) = 1-\hat{\lambda}_u [1 - H(x_p-u;\hat{\sigma}_u,\hat{\xi})]$,
$\hat{\lambda}_u$ is the proportion of the realisations of $X$ exceeding $u$ and $(\hat{\sigma}_u,\hat{\xi})$ are maximum likelihood estimates (MLEs)
obtained by using realisations of the threshold excesses. \citet{davison1990models} overview the properties of the GPD.

Threshold selection involves a bias-variance trade-off: too low a threshold is likely to violate the asymptotic basis of the GPD, leading to bias, whilst too high a threshold results in very few threshold excesses with which to fit the model, leading to large parameter and return level uncertainty. Thus, we must choose as low a threshold as possible subject to the GPD providing a reasonable fit to the data. There are a wide variety of methods aiming to tackle this problem
\citep{scarrott2012review,belzile2023} 
with the most commonly used methods suffering from subjectivity and 
sensitivity to tuning parameters. 

A novel automated approach to threshold selection is introduced by \citet{varty2021inference} specifically for modelling large, human-induced earthquakes. These data are complex due to improvements in measurement equipment over time. The major implication of such change is that data are missing-not-at-random, with the dataset appearing  to be realisations of a non-identically distributed variable, requiring a threshold $u(t)$ which varies with time $t$, even though the underlying process is believed to be identically distributed over $t$. Since excesses of $u(t)$ do not have the same GPD parameters over time, \citet{varty2021inference} transform these to a common standard exponential distribution via the probability integral transform, using estimates of $(u(t),\sigma_{u(t)},\xi)$. They then quantify the model fit using a metric based on a QQ-plot and select a time-varying threshold that optimizes this metric. The key novel aspect of their assessment is the use of bootstrapping methods in the metric evaluation which fully accounts for the uncertainty in the GPD fit, which varies across threshold choices.

Due to the lack of existing threshold selection methods designed for the context of \citet{varty2021inference}, that paper
focuses on the data analysis rather than investigating the performance of the threshold selection method.
We explore how their ideas can be best adapted to threshold selection in a
univariate, iid data context. We find that a variant of the \citet{varty2021inference} metric improves the performance and leads to substantially better results than existing automated methods, including 
greater stability  with respect to tuning parameters. 

We differ from \citet{varty2021inference} as we study both threshold selection and return level estimation when the truth is known. We also address an entirely different problem of how to incorporate the uncertainty resulting from threshold selection into return level estimation. Existing methods typically treat the threshold, once it has been selected, as known, for subsequent return level inference. The available data above candidate threshold choices are often few and so inference can be highly sensitive to the chosen threshold. Reliance on a single threshold leads to poor calibration of estimation uncertainty and as a result, can mislead inference. In particular, we show that the resulting confidence intervals for such an approach considerably under-estimate the intended coverage. We propose a novel and simple method, based on a double-bootstrap procedure, that incorporates the uncertainty in the selected threshold during inference. We show that the coverage probabilities of confidence intervals from our approach are close to the required nominal levels, thus ensuring our inferences provide meaningful information for design policies.

Ultimately, our aim is to provide a threshold selection method that does not require any user decisions to achieve adequate results. For example, the method should not be sensitive to the choice of candidate threshold grid, it should not require the estimation of a mode to select this grid, it should not have a limit on the number of candidate thresholds for a given sample size, nor should it exclude the possibility that the available data have been pre-processed, such as containing only the exceedances of some arbitrary level.

In Section~\ref{section: background}, we illustrate problems with threshold selection
and outline existing strategies.
Section~\ref{section: current}  describes the core existing automated methods while Section~\ref{section: ourmethod} introduces our procedure for 
the selection of a threshold, contrasting it with that of \cite{varty2021inference}. Section~\ref{section: uncertainty} presents our proposed method for incorporating threshold uncertainty into return level inference. In Section~\ref{section: simstudy}, the proposed methods are compared against existing methods on simulated data. In Section~\ref{section: resultsdata}, we apply our methodology to the widely studied troublesome dataset of the River Nidd, first analysed by \citet{davison1990models}.

\section{Background}\label{section: background}
The \textit{threshold stability property} of the GPD is key in many  threshold selection approaches: if excesses of a threshold $u$ follow a GPD then excesses of a higher threshold $v$ $(u < v < x^F)$ will also follow a GPD, with adjusted parameter values, i.e., if $X-u|(X>u) \sim \text{GPD}(\sigma_u,\xi)$, then $X-v|(X>v) \sim \text{GPD}(\sigma_u+\xi(v-u),\xi)$, see supplementary material \ref{section: thresh_stab}. By this property, the GPD shape parameter $\xi$ should have the same value for all valid choices of threshold. A modelling threshold can be selected as the lowest value for which this property holds, accounting for the sampling variability in the estimates of $\xi$. The conventional method for this assessment is known as a \textit{parameter stability plot} \citep{coles2001introduction}.  This plot displays the estimates of $\xi$ and their associated confidence intervals (CIs) for a set of candidate thresholds. The threshold is selected as the lowest value for which the estimate of $\xi$ for that level is consistent with estimates of $\xi$ at all higher thresholds. Throughout the paper, we use maximum likelihood estimation and parametric bootstrap-based CIs.

\begin{figure}[h!]
    \centering
    \includegraphics[width=0.95\textwidth]{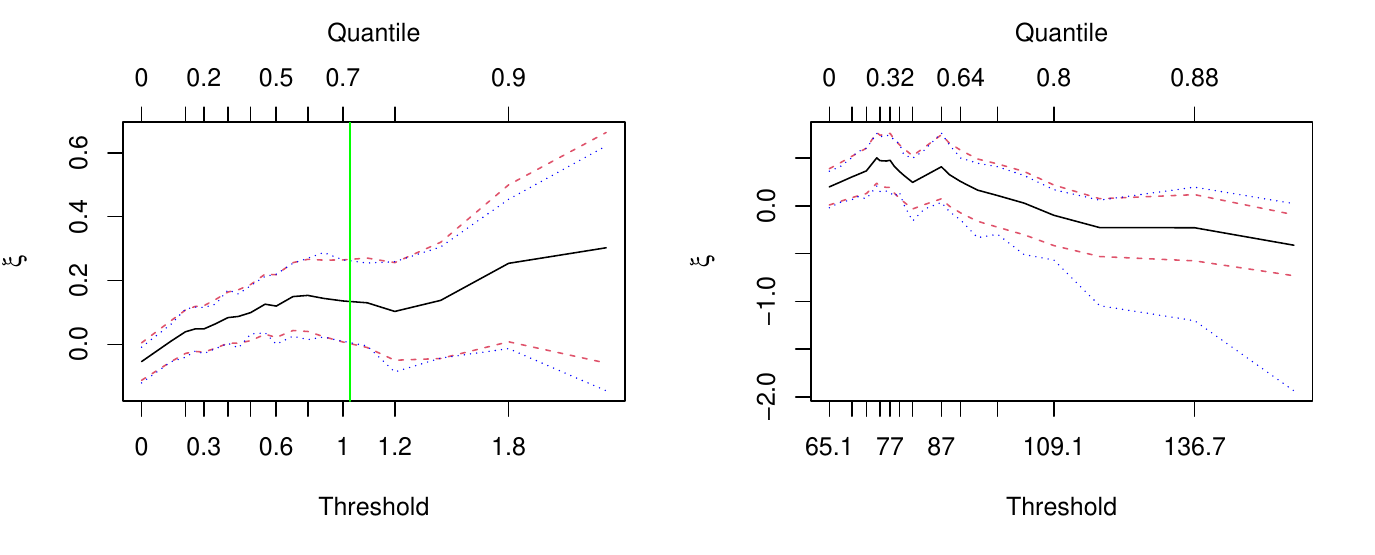}
    \caption{Examples of parameter stability plots with pointwise CIs using the delta-method [dashed] and bootstrapping [dotted] for [left] a simulated dataset with true threshold $u=1.0$ following Case 4 distribution [green-vertical] and [right] the River Nidd dataset.}
    \label{fig: parstabexamples}
\end{figure}

Figure~\ref{fig: parstabexamples} shows two parameter stability plots, with the left plot  for a simulated dataset of 1000 values generated from the Case 4 distribution, described in Section~\ref{section: simstudy}, where excesses of the threshold $u = 1.0$ follow a GPD(0.6, 0.1); and the right plot
for 154 measurements from the River Nidd. 
Each plot has  95\% CIs of two types; the delta method and the bootstrap. Profile log-likelihood based CIs were also evaluated but resulted in very similar intervals to the bootstrap method, so they were omitted. The delta method gives narrower CIs, though close to the bootstrap intervals for the larger dataset. Selecting an appropriate threshold using this method is challenging and subjective as the parameter estimates are dependent across threshold choices, there is a high level of uncertainty due to the small sample sizes that characterise extreme value analyses, and the uncertainty increases with threshold choice. 

For the Case~4 data, the plot shows 
that candidate thresholds above (below) $0.3$ are possibly appropriate (not appropriate) as CIs for higher candidate thresholds include (exclude) the corresponding shape parameter estimates, and above 0.8 the point estimates appear more stable. Here $(u, \xi)=(1, 0.1)$, so we can see that  candidates below 0.3 are not suitable as $\xi$ is outside their CIs, but the true threshold is higher than may be selected using this plot. For the River Nidd, lower candidate threshold values imply a very heavy-tailed distribution ($\hat{\xi} \approx 0.5$), whilst high candidate thresholds imply a very short tail, with estimates ($\hat{\xi} \approx -0.5$). 
As a result of this unusual behaviour, the Nidd data has become the primary example for non-trivial threshold selection \citep{davison1990models, northrop2014improved}. 
We apply our new method to this dataset in Section~\ref{section: resultsdata}. Further examples of the problems encountered when using parameter stability plots are given in supplementary material \ref{section: parstab}.

\citet{scarrott2012review}
and \citet{belzile2023}
review the extensive literature that aims to improve upon parameter stability plots. The latter characterises these methods, with a core reference, as follows: penultimate models \citep{northrop2014improved}, 
goodness-of fit diagnostics \citep{bader2018automated},
sequential-changepoint approaches
\citep{wadsworth2016exploiting}, predictive performance 
\citep{northrop2017cross}, and mixture models
\citep{naveau2016modeling}. It also discusses semi-parametric inferences \citep{danielsson2001using}, but it excludes the development by \cite{danielsson2019}, with similarities to the goodness-of-fit approaches. 

In Section~\ref{section: current}, we outline the key aspects of the core automated approaches with which we compare our proposed method. Supplementary material~\ref{section: NC_diag}
and \ref{section: SPDanielsson}
describe  \citet{northrop2014improved} 
and \citet{danielsson2001using,danielsson2019} respectively,
finding that the former suffers from subjectivity of interpretation similar to the parameter stability plots. We do not describe any mixture methods in this paper as although they benefit from accounting for threshold uncertainty, their inferences are strongly dependent on the choice of model for below the threshold, which we feel is inconsistent to the strategy of extreme value modelling and is likely to induce bias in the threshold selection and subsequent quantile estimation. 

\section{Existing automated threshold selection methods}\label{section: current}

Automated threshold selection methods aim to remove subjectivity from the choice of threshold by selecting an optimal threshold from a set of user-defined candidate thresholds based on optimising some criterion. We outline and compare the approaches of \citet{wadsworth2016exploiting} and \citet{northrop2017cross},
which we find to perform best of the considered existing methods. Further details of these methods are given in supplementary material~\ref{section: Scurrent}.

\citet{wadsworth2016exploiting} addresses the dependence between MLEs of $\xi$, denoted by $\hat{\bm{\xi}}$, over candidate thresholds. 
Using asymptotic theory for the joint distribution of  MLEs from overlapping samples of data, $\hat{\bm{\xi}}$ are transformed to the vector $\hat{\bm{\xi}}^*$ of normalised increments between successive $\hat{\bm{\xi}}$ values. For GPD data, asymptotically, $\hat{\bm{\xi}}^*$ would be iid realisations from a standard normal distribution, whereas if the data above any candidate threshold were not from a GPD, the associated
elements of $\hat{\bm{\xi}}^*$ would be better approximated by a non-standard normal. This changepoint behaviour is used to select the threshold. The underlying asymptotic arguments can cause considerable threshold sensitivity and the failure of the method to converge. Both issues are exacerbated by small samples and we identify systematic failures of the associated open source software when $\xi < 0$. To reduce such problems,
\citet[Table 1]{wadsworth2016exploiting} provides guidance on the number of candidate thresholds for a given sample size. 

\citet{northrop2017cross} model data using the  binomial-GPD (BGPD) 
model, which is GPD above $u$, with  $\lambda_u = \mathbb{P}(X>u)$ a model parameter, and an improper uniform density, of value $1-\lambda_u$, below $u$.  
They use Bayesian inference and, for each candidate threshold, assess the predictive density of GPD fits above a fixed validation threshold $v$, where $v$ is the largest candidate threshold.  
The selected threshold maximises the predictive ability of this model, above $v$, using leave-one-out cross-validation. The method is sensitive to the validation and candidate threshold set and to the prior joint density of the BGPD parameters.

\section{Novel metric-based constant threshold selection}\label{section: ourmethod}
\subsection{Metric choice}\label{subsec:metric}
We propose an adaptation of the \citet{varty2021inference} approach to identify the threshold $u$ for which the sample excesses, arising from iid and non-missing realisations of a continuous random variable, are most consistent with a GPD model. Both methods use a QQ-plot-based metric to approximate the integrated absolute error (IAE) between the quantiles of the model and the data-generating process. Our method, the \textit{expected quantile discrepancy} (EQD), uses the data on the original scale. In contrast, the method of \citet{varty2021inference} transforms the data to an Exponential$(1)$ marginal scale and will be termed the \textit{Varty method}. 

The following makes the difference between the two methods precise.  Let $\bm{x}_u = (x_1, \ldots , x_{n_u})$ be the sample of $n_u$ excesses of candidate threshold $u$ and $\bm{q} = \{q_i=(i-1)/(n_u-1): i= 1, \ldots, n_u\}$ be the vector of probability plotting points corresponding to the sample size of $\bm{x}_u$.
The sample quantile function $Q(\cdot; \bm{x}_u, \bm{q}) : [0,1] \rightarrow \mathbb{R}^+$ is defined as the linear interpolations of the points $\left\{(q_i, {x}^{(i)}_u): i = 1,\ldots,n_u\right\}$, with
$x^{(i)}_u$ denoting the $i^{\text{th}}$ order statistic of $\bm{x}_u$ (increasing with $i$), where any ties are handled similarly through linear interpolation. The transformation to Exponential(1) margins is defined by $T(x;\sigma,\xi) = F^{-1}_{\text{Exp}}\{H(x;\sigma, \xi)\}$ where $F^{-1}_{\text{Exp}}$ is the inverse distribution function of an Exponential(1) variable, $H$ is the GPD function~\eqref{eqn: gpd}, and let $T(\bm{x}_u;\sigma_u,\xi) =\{T(x_1;\sigma_u, \xi), \ldots,T(x_{n_u};\sigma_u, \xi)\}$. To incorporate the effect of sampling variability in the data into the threshold choice, the expected (average) deviation over the QQ-plot, calculated for the probabilities $\{p_j = j / (m+1): j = 1,\dots,m\}$, is calculated across bootstrapped samples of $\bm{x}_u$, denoted $\bm{x}^b_u$ for the $b^{\text{th}}$ bootstrap sample, $b=1, \ldots ,B$. For both methods, this results in the overall measure of fit 
$\hat{d}_E(u) = \sum_{b =1}^{B} d_b(u)/B$, where 
\begin{equation}
    d_b(u) = 
    \begin{cases}
        \frac{1}{m} \sum\limits_{j=1}^{m} \mid \frac{\hat{\sigma}^b_u}{\hat{\xi}^b_u}[(1-p_j)^{-\hat{\xi}^b_u}-1] - Q(p_j; \bm{x}^b_u, \bm{q})\mid & \text{EQD}\\
        \frac{1}{m} \sum\limits_{j=1}^{m} \mid-\log(1-p_j) - Q(p_j; \hat{T}(\bm{x}^b_u; \hat{\sigma}^b_u, \hat{\xi}^b_u), \bm{q})\mid & \text{Varty},
    \end{cases}
    \label{eqn: d_i}
\end{equation}
and $(\hat{\sigma}^b_u, \hat{\xi}^b_u)$ are the estimated GPD parameters fitted to the bootstrapped sample $\bm{x}^b_u$.  The selected threshold minimises the estimated IAE, $\hat{d}_E$, over a set of candidate thresholds. In Sections~\ref{section:whyEQDchoice} and \ref{subsec:tuning}
respectively, 
we justify the choices made in the formulation of the EQD metric and discuss our recommendation for default values
for the tuning parameters $(B,m)$.

In supplementary material \ref{section: comp_varty}, we compare the EQD and Varty methods through an extensive simulation study to assess which version of
metric~\eqref{eqn: d_i} performs better for threshold selection and quantile estimation. 
For threshold selection, the methods perform similarly; each 
method narrowly achieves the smallest root-mean-square error (RMSE) in two of Cases 1-4, discussed in Section~\ref{section: simstudy}.
However, for the estimation of high quantiles, the EQD outperforms the Varty method obtaining the lowest RMSE in the majority of cases and quantiles, due to  having the smaller variance of estimates. We ultimately aim to estimate high quantiles accurately following threshold selection. Given that this study indicates that the EQD should be preferred for this aim and to avoid unnecessary repetition, we omit the results for the Varty method for the remainder of the paper.

\subsection{Investigation of the EQD metric choice}
\label{section:whyEQDchoice}
For a given $u$, $d_b(u)$ evaluates the mean-absolute deviation between the $b^{\text{th}}$ bootstrap sample quantiles and the fitted model-based quantiles, i.e., the mean-absolute deviation from the line of equality in a QQ-plot for that particular bootstrap sample.  
This type of assessment by itself is not radical, 
as for any observed sample data, QQ-plots are the standard method of assessing model fit \citep{coles2001introduction}. The novelty for assessing the validity of a candidate threshold $u$ comes from the way that the EQD metric is constructed.

There are 
a number of novel choices which we have made in the EQD metric that require justification, in particular; the use of the mean-absolute deviation;
the choice of quantiles and their interpolation in the QQ-plot; the use of bootstrap samples; and that the observed data are not explicitly used in the metric. We examine each of these features in the supplementary material, through simulation studies involving the case studies of 
Section~\ref{section: simstudy}. For each feature, we find positive evidence for our selections. Below, we explain why we made these choices and outline how they performed relative to other alternative formulations.

We focus on the mean-absolute deviation on the QQ scale as \citet{varty2021inference} found that this was more effective than using the mean-squared deviation on that scale and either metric on the PP scale. Our simulation studies also found this to be a more robust measure of fit than the maximum deviation, as proposed by \citet{danielsson2019}.  

We choose to take $\{p_j\}$ to be equally-spaced and to weight contributions to $d_b(u)$ equally across the corresponding quantiles. Although higher (lower) sample quantiles exhibit greater (less) sampling variability, equal weighting is appropriate when taking the $\{p_j\}$ values to be equally-spaced because for any $\xi>-1$, the GPD density is monotonically decreasing. This leads to dense evaluation for lower sample quantiles and more sparse evaluation in the upper tail. The choice of equal weighting on this scale is motivated and supported by empirical evidence in \citet{varty2021inference}. Our choice for $p_j$ is based on the expression for plotting points in a QQ-plot assessment we use in previous research and the choice for $q_i$ is the default option for the R \verb|quantile| function. As these choices are subjective, we also consider alternative definitions but find that there is no systematic ordering of the performance over these definitions and any differences in RMSE for the thresholds selected by the EQD are minimal, especially when compared to the differences between the EQD and existing methods. 

The average over bootstrapped samples in metric~\eqref{eqn: d_i} is not a standard use of bootstrapping, i.e., we utilise the bootstraps in a measure of fit rather than to describe the uncertainty in some estimated quantity. Our aim in doing this is to account for the sampling variability in the observed data, thus avoiding over-fitting of the GPD model to the observed dataset which could lead to higher threshold choices than necessary, reduced numbers of exceedances, and extra uncertainty in parameter and quantile estimates. To confirm this, we considered using only the observed sample values in the metric. This leads to higher and more variable thresholds choices in a variety of cases and an overall performance which is either noticeably worse or at best, comparable to our approach.

One may also be concerned that $\bm{x}_u$ is not included directly in metric~\eqref{eqn: d_i}. To address this, 
we additionally explored the effect of using $Q(p_j; \bm{x}_u, \bm{q})$ instead of $Q(p_j; \bm{x}^b_u, \bm{q})$ within the EQD metric, despite it being unconventional to compare sample quantiles to those of a model fitted using a different (bootstrapped) sample. We found no benefit to doing so. Moreover, using only $\bm{x}_u$ to estimate the IAE ignores that this estimate would change for another realisation of the same data generating process and that variability in this estimate increases with $u$. Our approach utilises the bootstrap resamples in the measure of fit to provide more stability in the threshold choice and allow us to account for the increasingly uncertain parameter and quantile estimates as the threshold increases.

\subsection{Choice of tuning parameters}
\label{subsec:tuning}

An in-depth study in supplementary material~\ref{section: sensitivity_analysis} demonstrates that the EQD method is robust to the choice of the tuning parameters $B$ and $m$. Consequently,  we take $(B,m)=(100,500)$ throughout the paper and in the supplementary material, unless stated otherwise.

The number of bootstrapped samples $B$  controls the level of sampling variability that is incorporated into the threshold choice and so we expect higher values of $B$ to lead to more stable threshold choices. The RMSE values for threshold estimation reflect this but also show that computation time increases linearly with $B$. 
For a one-off analysis, there is certainly merit in taking as large a value for $B$ that is computationally feasible. For simulation studies, when the computational implications of the choice of $B$ are more important, we find that $B=100$ balances accuracy and computation time sufficiently. 

The tuning parameter $m$ gives the number of equally-spaced evaluation probabilities used in expression~\eqref{eqn: d_i}. 
The EQD metric aims to approximate the IAE between model
quantiles and quantiles of the data generating process (i.e., not for a particular sample) and a larger choice of $m$ improves this approximation. To compare fairly across a range of candidate thresholds, we choose to keep the quality of the approximation of the IAE fixed across thresholds and bootstraps by fixing the number $m$ of points in the quantile interpolation grid. 

For a particular bootstrap sample, this choice of fixed $m$ can lead to under- or over-sampling of the upper tail depending on whether $m < n_u$ or $m > n_u$. We explore the sensitivity of the EQD method to $m$ with $m=cn$ and $m=cn_u$, with $c=0.5,1,2,10$. For both strategies, we find that increasing $m$ beyond 500 essentially wastes the increased computation time as the RMSE values for threshold estimates showed little sensitivity to $m$. We also explore the effect of the interpolation grid on the sampling distribution of $d_b(u)$ values, over different thresholds, when evaluated using $m=500$ or $m=n_u$.
We find that there is little effect from the choice of interpolation grid outside of the very highest candidate thresholds, but these differences have no effect on the selected threshold in our examples.
We conclude that $m=500$, is suitable as a default value in practice but we can see merits in also ensuring that $m \geq \underset{u}{\max}(n_u)$, where the maximisation is across all candidate thresholds.

\section{Accounting for parameter and threshold uncertainty}\label{section: uncertainty}

Even if the true threshold $u$ is known, relying on point estimates for the GPD parameters results in misleading inference \citep{coles2003anticipating}. CIs 
are needed, but as standard errors and profile likelihoods rely on asymptotic arguments, they are not ideal due to the sparsity of threshold exceedances.
We prefer parametric bootstrap methods which, as discussed in Section~\ref{section: background}, perform similarly to the profile likelihood
for large samples. Algorithm 1 details the bootstrapping procedure to account for GPD parameter uncertainty when $u$ is known. A GPD is fitted to the $n_u$ data excesses of $u$ from a sample $\bm{x}$ of size $n$ $(n \geq n_u)$. Using the fitted parameters, $B_1$ parametric bootstrap samples above $u$ are simulated, each of size $n_u$, and the GPD is re-estimated for each sample. A summary statistic, e.g., a return level, $s(u, \lambda_u, \sigma_u, \xi)$, may be computed for each of the $B_1$ bootstrapped values for $(\sigma_u, \xi)$. This enables the construction of CIs for the GPD parameters and return levels.

\begin{algorithm}
\caption{Parameter uncertainty for known threshold}\label{alg: parameter_uncertainty}
\begin{algorithmic}
\Require $(\bm{x},u, B_1)$
\State Find $n_u=\#\{i: x_i>u\}$, set 
$\hat{\lambda}_u = n_u/n$, and fit a GPD to $\bm{x}$ data above $u$ to give $(\hat{\sigma}_u, \hat{\xi}_u)$.  
\For{$b=1, \ldots, B_1$}
\State Simulate sample $\bm{y}^b_u$ consisting of $n_u$ excesses of $u$ from GPD$(\hat{\sigma}_{u}, \hat{\xi}_u)$.
\State{Obtain parameter estimates $(\hat{\sigma}_b, \hat{\xi}_b)$ for $\bm{y}^b_u$ and summary of interest $s(u,\hat{\lambda}_u,\hat{\sigma}_b, \hat{\xi}_b)$.}
\EndFor \\
\Return A set of $B_1$ bootstrapped estimates for the summary statistic of interest.
\end{algorithmic}
\end{algorithm}

Algorithm 1 focuses on the uncertainty of the estimates of $(\sigma_u, \xi)$. We incorporate the additional uncertainty in the estimation of $\lambda_u$ by replacing the fixed $n_u$ in the loop over $b$ with a random variate from a Bin$(n, \hat{\lambda}_u)$ distribution for each bootstrap sample, with this extension then referred to as Algorithm 1b.


GPD inferences are sensitive to the choice of threshold \citep{davison1990models} but uncertainty about this choice is not represented in Algorithms 1 or 1b. This omission would be important when return levels inform the design of hazard protection mechanisms, where omitting this source of uncertainty could lead to over-confidence in the inference and have dangerous consequences. Algorithm 2 provides a novel method to propagate both threshold and parameter uncertainty through to return level estimation, using a double-bootstrap procedure. To focus on the threshold uncertainty and to forgo the need for a parametric model below the threshold, we employ a non-parametric bootstrap procedure on the original dataset. We resample with replacement $n$ values from the observed data $B_2$ times, estimate a threshold for each such bootstrap sample using the automated selection method of Section~\ref{section: ourmethod}, and fit a GPD to the excesses of this threshold. For each one of the $B_2$ samples, we employ Algorithm 1 to account for the subsequent uncertainty in the GPD parameters. Calculating a summary statistic for each of the $B_1 \times B_2$ samples leads to a distribution of bootstrapped estimates that accounts for uncertainty in the threshold selection as well as in the GPD and threshold exceedance rate parameters. We use $B_1=B_2=200$. To run this algorithm using the EQD method for the threshold selection step (which itself has $B$ bootstraps), it would require $B_2 (B + B_1)$ bootstrap samples to be generated. Specifically, for the $B_2$ samples initially generated for Algorithm 2, we have $B_2\times B$ in selecting the threshold values and $B_2 \times B_1$ in capturing the GPD parameter uncertainty above these selected thresholds. this can res
In Section~\ref{section: simstudy}, we illustrate how using Algorithm~2 improves the coverage probability of CIs, and in Section~\ref{section: resultsdata} how it widens  CIs for return levels of the River Nidd. 

\begin{algorithm}
\caption{Parameter uncertainty for unknown threshold}\label{alg: thresh_par_uncertainty}

\begin{algorithmic}
\Require $(\bm{x}, n, B_2, B_1)$
\For{$b=1,\ldots,B_2$}
\State Obtain sample $\bm{x}_b$ of size $n$ by sampling $n$ times with replacement from $\bm{x}$.
\State Estimate threshold $\hat{u}_b$ for $\bm{x}_b$ and record number of excesses as $n_{\hat{u}_b}$.
\State Employ Algorithm 1 with inputs: $(\bm{x}_b, \hat{u}_b, B_1)$.
\EndFor \\
\Return A set of $B_1 \times B_2$ bootstrapped estimates for the summary statistic of interest.
\end{algorithmic}
\end{algorithm}

\section{Simulation study}\label{section: simstudy}
\subsection{Overview}\label{section: simstudyoverview}
We illustrate the performance of the EQD method against the \citet{wadsworth2016exploiting} and \citet{northrop2017cross} approaches, 
which we term the \textit{Wadsworth} and \textit{Northrop} methods respectively. \citet{danielsson2001using,danielsson2019} approaches perform considerably worse than all others in threshold selection and quantile estimation; so results for these methods are only given in supplementary material \ref{section: detailed_sim_res}. We utilised the following R code for Wadsworth, Northrop and EQD methods respectively: code given in the \href{https://www.tandfonline.com/doi/full/10.1080/00401706.2014.998345}{supplementary materials} of \citet{wadsworth2016exploiting}, \textit{threshr} \citep{threshr}, and \if0\blind{\url{https://github.com/conor-murphy4/automated_threshold_selection} \citep{Murphy_Software_for_threshold_2023}}\fi.

The performance of all of the methods depends somewhat on the choice of the set of candidate thresholds which we denote by:
\begin{equation}
    C_u=\{u_i, i=1,\ldots ,k: u_1< \ldots <u_k\},
    \label{eqn: candidate_thresh}
\end{equation}
where we restrict the $u_i$ to be sample quantiles evaluated at equally-spaced probabilities. The range $[u_1,u_k]$, the number of candidates $k$ and the inter-threshold spacing are all potentially important in terms of how they affect the performance of the methods. As emphasised in Section~\ref{section: intro}, we are aiming for an automated threshold selection method which can achieve accurate results without any user inputs, so a key element of our study is to investigate how these features of the set $C_u$ impact on the methods' relative performance.  When fitting a GPD with decreasing density (i.e., for $\xi > -1$), it would be inadvisable to use a threshold which clearly lies below the mode of the distribution. As we want to avoid the requirement of user estimation of the mode, our standard choice for the range of the candidate grid is $[u_1,u_k]$: $(u_1,u_k)=(0\%,95\%)$ sample quantiles of all the data. However, we also explore several cases
where only the data lying above the mode
are used  with $[u_1,u_k]$: $(u_1,u_k)=(0\%,95\%)$ now sample quantiles of the remaining data. To remove the uncertainty arising from the choice of estimator of the mode, we use the true mode which has a unique value in our simulated cases. Results in supplementary material~\ref{section: mode_candidate_grids} indicate that our original choice for the candidate threshold set does not unfairly favour the EQD method in any way. 

We consider two scenarios: Scenario~1 and Scenario~2 where the true threshold is known and unknown respectively. Here, we present the results using a candidate
threshold grid across the whole distribution for Scenario~1 and above the sample median for Scenario~2, with the latter chosen as the Wadsworth method fails when applied across the default range in that setting. The Wadsworth method relies on asymptotic arguments, which limits how large $k$ can be relative to the sample size, $n^{\prime}$, above the mode, with $n^{\prime} \leq n$, where $n$ is the total sample size. To assess how the Wadsworth method performs as a fully automated method, we apply the method despite the value of $k$ not always aligning with the guidance in \citet{wadsworth2016exploiting} about its size relative to $n^{\prime}$.

We assess the methods' ability to estimate the true threshold (when it exists) and  the true return levels, using the RMSE to measure performance. The true quantiles and  all bias-variance components of RMSE, discussed in this section, are given in supplementary materials~\ref{section: case_dist} and \ref{section: detailed_sim_res} respectively. We also investigate the merits of including the uncertainty in threshold selection in our inference, as discussed in Section~\ref{section: uncertainty}, 
in terms of how they improve the coverage levels of CIs relative to their nominal values.

\subsection{Scenario 1: True threshold for GPD tail}
\label{subsec:Scenario1}
We consider Cases 0-8, with different properties above and below the true threshold of $u=1.0$ and various sample sizes. Case~0, where all of the data are from a GPD, is reported in supplementary material \ref{section: furthercases}, with the EQD performing notably better than the existing methods. Here, we present detailed results for Cases 1-4, with Table~\ref{tab: case_descriptions} providing outline model and sample size properties, with full details and density plots given in  supplementary material~\ref{section: case_dist}. Cases 5-8 are considered briefly after discussing Cases 1-4 below.

Cases 1-3 all have a distinct changepoint in the density and density mode both at the true threshold which should make all methods of threshold selection perform better than in situations without either of these features. Cases 1 and 2 have the same distribution, with $\xi>0$, with Case 2 having a smaller sample size. We find that the Wadsworth method fails to estimate a threshold in samples with $\xi < -0.05$ irrespective of sample size, so 
Case 3 is selected near that boundary where the method works reasonably and has double the sample size of Case 1. 
Case 4 provides a more difficult example with a continuous density and a small number of exceedances of the true threshold. The data are derived from a partially observed GPD, denoted GPD\textsubscript{p}, with data drawn from a GPD above 0 and rejected if less than an independent realisation from a Beta(1,2) distribution. 

For each case, the results are based on 500 replicated samples, for which  we test the candidate thresholds $C_u$, with $k=20$, as given in \eqref{eqn: candidate_thresh}, with the true threshold being the $16.67\%$ quantile for Cases 1-3 and the $72.10\%$ quantile for Case 4.
\begin{table}[h!]
    \centering
    \begin{tabular}{|c||c|c||c|c|}
        \hline
        Models & Below threshold & Sample size & Above threshold & Sample size \\
        \hline
        Case 1 & U$(0.5, 1.0)$ & 200  & GPD$(0.5, 0.1)$ & 1000  \\
        Case 2 & U$(0.5, 1.0)$ & 80  & GPD$(0.5, 0.1)$ & 400  \\
        Case 3 & U$(0.5, 1.0)$ & 400  & GPD$(0.5, -0.05)$ & 2000  \\
        Case 4 & GPD\textsubscript{p}$(0.5, 0.1)$ & 721  & GPD$(0.6, 0.1)$ & 279\\
        \hline
    \end{tabular}
    \caption{Model specifications for Cases 1-4.}
    \label{tab: case_descriptions}
\end{table}

\FloatBarrier

\textbf{Cases 1-4, Threshold recovery:}
Table~\ref{tab: rmse_thr} shows the RMSE of the chosen thresholds for each method in Cases 1-4, with the EQD achieving RMSEs 1.2-7.7 (1-11.2) times smaller than the Wadsworth (Northrop) method. The EQD has the lowest bias by a considerable margin in Cases 1-3 and shows the lowest variance in threshold estimation in all cases. In fact, the variance is reduced by a factor of at least 20 relative to both the Wadsworth and Northrop methods (see Table~\ref{tab: biasvarcases}). The very strong performance of the EQD relative to both the Wadsworth and Northrop methods is particularly noteworthy in Cases 1-3, and is also seen for Case 0 and later for Cases 5-7. We believe that the key reason for this is the discontinuity in the density, a feature common to all of these cases, as that appears to lead to a very small bias for the EQD method relative to the other methods. Specifically, the variance penalty  of the EQD metric seems to push the threshold as low as possible, but its complementary goodness-of-fit measure almost entirely stops the threshold being selected below the clear discontinuity in the density. For Case 4, which has a continuous density, the EQD achieves the smallest RMSE almost entirely due to it having the smallest variance but with a bias component broadly comparable with the other methods.

\begin{table}[ht!]
    \centering
    \begin{tabular}{|c||c||c||c|}
        \hline
        & \textit{EQD} & \textit{Wadsworth\tablefootnote{Results for Wadsworth are calculated only on the samples where a threshold was estimated. It failed estimate a threshold for 2.4\%, 26.4\%, 0\%, 4.4\% of the simulated samples in Cases 1-4, respectively. \label{note_wads_fail_1}}} & \textit{Northrop}\\
        \hline
        Case 1 & $\bm{0.048}$  & 0.349  & 0.536  \\
        Case 2 & $\bm{0.060}$  & 0.461  & 0.507  \\
        Case 3 & $\bm{0.060}$  & 0.221  & 0.463  \\
        Case 4 & $\bm{0.526}$  & 0.628  & 0.543  \\
        \hline
    \end{tabular}
    \caption{RMSE of the threshold choices for each method-case combination. The smallest values for each case are highlighted in bold.}
    \label{tab: rmse_thr}
\end{table}

\textbf{Cases 1-4, Quantile recovery:}
Table~\ref{tab: rmse_quant} presents the RMSEs for the $(1-p_{j,n})$-quantiles where $p_{j,n} = 1/(10^jn)$ for $j=0,1,2$ for sample size $n$, which ensures that extrapolation is equally difficult over $n$ for a given $j$. When $j=0$, no extrapolation is required so the choice of $u$ should not be too important; the similar RMSEs across methods reflect this. As $j$ increases, all RMSEs increase and the differences between methods become clear. The EQD method is best uniformly, followed by the Wadsworth and then the Northrop methods. This pattern reflects the findings in Table~\ref{tab: rmse_thr}, although here with  differential performances sensitive to $j$. However, in terms of quantile estimation, the EQD method does not retain the large differential relative to the other methods which was seen for threshold selection in Cases 1-3. 
In contrast, the differences between methods in Case 4 are now more apparent, as controlling the variance is more important than any small differences in bias when we are concerned with a RMSE assessment of quantiles which lie far into the tail.
The EQD achieves the lowest bias in the majority of cases and leads to quantile estimates with considerably less variance in all cases, particularly as $j$ increases.


\begin{table}[ht!]
    \centering
    \begin{tabular}{ |c|c|c|c||c|c|c| } 
    \hline 
     & \textit{EQD} & \textit{Wadsworth\footref{note_wads_fail_1}} & \textit{Northrop} &  \textit{EQD} & \textit{Wadsworth\footref{note_wads_fail_1}} & \textit{Northrop} \\
     \hline 
     $j$ & \multicolumn{3}{|c||}{\textbf{Case 1}} & \multicolumn{3}{|c|}{\textbf{Case 2}} \\
     \hline
     0 & $\bm{0.563}$ & 0.594  &  0.755 &  $\bm{0.599}$ & 0.631 & 0.736 \\ 
     1 & $\bm{1.258}$ & 1.391 & 2.376 &  $\bm{1.488}$ & 1.644 & 3.513 \\ 
     2 & $\bm{2.447}$ & 2.717 & 7.097 &  $\bm{3.119}$ & 3.484  &  22.916 \\       
     \hline
     \hline
    & \multicolumn{3}{|c||}{\textbf{Case 3}} & \multicolumn{3}{|c|}{\textbf{Case 4}} \\
     \hline
     0 & $\bm{0.190}$ & 0.195  &  0.230 & $\bm{0.677}$ & 0.800 & 0.791 \\ 
     1 & $\bm{0.323}$ & 0.344 & 0.450 &  $\bm{1.563}$ & 2.059 & 2.217 \\ 
     2 & $\bm{0.483}$ & 0.516 & 0.744 &  $\bm{3.043}$ & 4.485 & 5.568 \\       
     \hline
    \end{tabular}
\caption{RMSEs in the estimated quantiles in Cases 1-4 based on fitted GPD above chosen threshold. The smallest RMSE for each quantile are highlighted in bold.}
 \label{tab: rmse_quant}
\end{table}

\textbf{Summary for Cases 5-8:} Cases 5-8 are very similar in form to Cases 1-3 but with different shape parameters and sample sizes. The results for these cases are presented in supplementary material \ref{section: furthercases}, with a brief summary given here. Specifically, for Cases 5-7, we find that the EQD exhibits the strongest performance and the Wadsworth method consistently fails due to the small sample sizes or computational issues with numerical integration when $\xi < -0.05$. Case~8 is parameterised similarly to Case 1 but with an unrealistic sample size of $n=20000$. Although the data in Case~8 are more suited to a method reliant on asymptotic theory, the EQD performs comparably with the Wadsworth method, with both performing better than the Northrop method.

\textbf{Case 4, True quantile coverage:}
We apply Algorithms 1, 1b and 2 to data from Case~4, the hardest case for threshold selection. Table~\ref{tab: coverage_case4} presents the coverage probabilities of the nominal 80\% and 95\% CIs of the estimated $(1-p_{j,n})$-quantiles as well as the average ratio of the CI widths (based on \textit{Alg 2} relative to \textit{Alg 1}) over the 500 samples, termed CI ratio. Results for extra quantile levels, as well as coverage for the 50\% CI, are given in supplementary material \ref{section: further_coverage}. Overall, incorporating only parameter uncertainty (\textit{Alg 1} and \textit{Alg 1b}) leads to underestimation of interval widths and inadequate coverage of the true quantiles, especially as we extrapolate further. The additional uncertainty, given in 
\textit{Alg~1b}, by also accounting for uncertainty in the rate of threshold exceedance, typically makes a very small improvement in coverage, and for some quantiles, this actually leads to a reduction in coverage due to Monte Carlo variation in the simulations.
In contrast, the inclusion of the additional threshold uncertainty (\textit{Alg 2}) leads to much more accurate coverage of the true quantiles 
across all exceedance probabilities.
The CI ratios show that this highly desirable coverage is achieved with only 43-62\% increase in the CI widths on average.
\begin{table}[h!]
    \centering
    \begin{tabular}{|c|c|c|c|c|c|c|}
     \hline
     & \multicolumn{3}{|c|}{80\% confidence} & \multicolumn{3}{|c|}{95\% confidence}\\
     \hline
     $j$ & 0 & 1 & 2 & 0 & 1 & 2 \\
     \hline
     \textit{Alg 1} & 0.646 & 0.618  & 0.606 & 0.834  & 0.804  & 0.794 \\
     \textit{Alg 1b} & 0.656 & 0.638 & 0.612 & 0.830 & 0.814 & 0.794 \\
     \textit{Alg 2} & 0.798  & 0.772 & 0.758 & 0.954  & 0.948  & 0.944 \\
     CI ratio &  1.430 & 1.452  & 1.475 & 1.484 & 1.546 & 1.621 \\
     \hline
    \end{tabular}
    \caption{Coverage probabilities for estimated quantiles using Algorithms 1, 1b and 2 for 500 replicated samples from Case 4 with sample size of 1000. CI ratio gives the average ratio of the CIs for Algorithm 2 relative to Algorithm 1 over the 500 samples.}
    \label{tab: coverage_case4}
\end{table}

\vspace{-0.5cm}
\subsection{Scenario 2: Gaussian data}
In applications, there is no true or known value for the threshold
above which excesses follow a GPD, so we explore this case here. We select the standard Gaussian distribution as it has very slow convergence towards an extreme value limit \citep{gomes1994penultimate}, so threshold selection is likely to be difficult. We assess threshold selection methods based on estimation of the true quantiles 
$\Phi^{-1}(1-p_{j,n})$ where $p_{j,n}=1/(10^jn)$, for $j=0,1,2$.
We simulate 500 samples, for $n=2000$ and $20000$, 
with $C_u$, given in \eqref{eqn: candidate_thresh}, now having range $[u_1,u_k]:(u_1,u_k) = (50\%,95\%)$ sample quantiles of the data and $k=10$ and 91 (i.e., steps of 5\% and 0.5\%) for the two choices of $n$ respectively. As with Case~8 in Section~\ref{subsec:Scenario1}, $n=20000$ is unrealistic, but we include it to illustrate the slow convergence. 

\textbf{Quantile recovery:}
Table~\ref{tab: rmse_norm} shows the RMSEs of the estimated quantiles. For $n=2000$, the EQD method achieves the smallest RMSE with the Northrop method a close second, with the reverse when $n=20000$. 
The median and 95\% CI of the chosen thresholds are given in supplementary material \ref{section: detailed_sim_res}. The Northrop method tends to choose slightly higher thresholds than the EQD method, leading to a small reduction in bias, but for only the smaller $n$ is the additional variability relative to the EQD a disadvantage. The Wadsworth method performs the worst, selecting lower thresholds and so incurring the most bias. 
\begin{table}[ht]
    \centering
    \begin{tabular}{|c|c|c|c||c|c|c|}
    \hline
    & \multicolumn{3}{|c||}{$n=2000$} & \multicolumn{3}{|c|}{$n=20000$} \\
    \hline
     $j$ & \textit{EQD} & \textit{Wadsworth}\tablefootnote{Results for the Wadsworth method, which failed on 0.4\% of the samples here, are calculated only for samples where a threshold estimate was obtained. \label{note_wads_fail_2}} & \textit{Northrop} & \textit{EQD} & \textit{Wadsworth} & \textit{Northrop} \\
     \hline
     0 & $\bm{0.214}$ & 0.239 & 0.225 & 0.187 & 0.214 & $\bm{0.172}$ \\
     1 & $\bm{0.430}$ & 0.529 & 0.461 & 0.368 & 0.422 & $\bm{0.331}$ \\
     2 & $\bm{0.703}$ & 0.890 & 0.765 & 0.594 & 0.672 & $\bm{0.533}$ \\
     \hline
    \end{tabular}
    \caption{RMSEs of estimated $(1-p_{j,n})$-quantiles for 500 replicated samples from a Gaussian distribution for samples of size $n$. The smallest RMSE are highlighted in bold.}
    \label{tab: rmse_norm}
\end{table}

\vspace{-0.3cm}
\textbf{True quantile coverage:}
For assessing the coverage of true quantiles using Algorithms 1, 1b and 2 for Gaussian data, Table~\ref{tab: coverage_norm} presents the coverage probabilities of the nominal 80\% and 95\% CIs of the estimated quantiles, when $n=2000$, as well as the average ratio of the CI widths (again, of \textit{Alg 2} relative to \textit{Alg 1}) over the 500 samples, with more results given in supplementary material \ref{section: further_coverage}. Across the $p_{j,n}$, both \textit{Alg 1 and 1b} give very low coverage probabilities in both cases, with performance deteriorating as $j$ increases. The added threshold uncertainty from \textit{Alg 2} results in large increases in coverage though still somewhat less than required, with this achieved through increases in CI widths by 45-66\% on average.  This weaker performance than we find in
Section~\ref{subsec:Scenario1} suggests that no sample threshold (for realistic sample sizes) is large enough to overcome bias in making extreme value approximations for Gaussian data, but the improvement in coverage using \textit{Alg~2} demonstrates the importance of including the additional threshold uncertainty.
\begin{table}[ht]
    \centering
    \begin{tabular}{|c|c|c|c|c|c|c|}
     \hline
     & \multicolumn{3}{|c|}{80\% confidence} & \multicolumn{3}{|c|}{95\% confidence}\\
     \hline
       $j$ & 0 & 1 & 2 & 0 & 1 & 2 \\
       \hline
     \textit{Alg 1} & 0.588  & 0.450 &  0.366 & 0.750 & 0.618 &  0.510 \\
     \textit{Alg 1b} & 0.592  & 0.442 &  0.364 & 0.746 & 0.620 &  0.508 \\
     \textit{Alg 2} & 0.718 & 0.598 &  0.492  & 0.866  & 0.814 &  0.756  \\
     CI ratio & 1.457 & 1.480 &  1.509 & 1.495 & 1.576 & 1.665  \\
      \hline
    \end{tabular}
    \caption{Coverage probabilities for estimated quantiles using Algorithms 1, 1b and 2 for 500 replicated samples from a Gaussian distribution with sample size of 2000. CI ratio gives the average ratio of the CIs for Algorithm 2 relative to Algorithm 1 over the 500 samples.}
    \label{tab: coverage_norm}
\end{table}

\section{Application to river flow data}\label{section: resultsdata}
The River Nidd dataset consists of 154 storm event peak daily river flow rates that exceeded 65 m$^3$/s in the period 1934-1969, i.e., an average exceedance rate of 4.4 events per year. Each observation can be deemed ``extreme" and iid, though not necessarily well-described by a GPD. \citet{davison1990models} identify the difficulties these data present for threshold selection and parameter uncertainty, which we reiterated in discussion of Figure~\ref{fig: parstabexamples}. Given the small sample size for the River Nidd, 
any increase in the threshold value is more significant in terms of parameter uncertainty, than for larger datasets studied in Section~\ref{section: simstudy}.

Table~\ref{tab: Nidd_grids} shows the selected thresholds of each of the methods for a range of candidate grids\footnote{In marked cases, the Northrop method outputted a chosen threshold with some convergence warnings. \label{convergence}}. The remarkable robustness of the EQD (evaluated with $B=200$ bootstrap samples)
across grids stems from the method's novel incorporation of data uncertainty. The Wadsworth method fails to estimate a threshold unless the  grid is made very coarse, and even then exhibits considerable sensitivity (varying between 0\% and 90\% sample quantiles) over grids of equal size but different endpoints and increments. This is problematic as a coarse grid is likely to remove the most appropriate threshold from consideration.
The Northrop method critically depends on the validation threshold, and we find that increasing this level above the $90\%$-quantile leads to failure or convergence warnings. The thresholds selected by this method are quite variable 
(between 0\% and 80\% sample quantiles) over the grids.

\begin{table}[h]
    \centering
    \begin{tabular}{|c|c|c|c|}
    \hline
    \multicolumn{4}{|c|}{Estimated thresholds for the River Nidd dataset} \\
    \hline
     Grid (\% quantile) &  \textit{EQD} & \textit{Wadsworth} & \textit{Northrop}\\ 
     \hline
      0 (1) 93 & 67.10 (3\%) & NA & 68.45\footref{convergence} (6\%) \\
      0 (1) 90 & 67.10 (3\%) & NA & 65.08 (0\%) \\ 
      0 (1) 80 & 67.10 (3\%) & NA & 100.28 (75\%) \\ 
      0 (20) 80 & 65.08 (0\%) & NA & 109.08 (80\%) \\
      0 (30) 90 & 65.08 (0\%) & 149.10 (90\%) & 65.08 (0\%) \\ 
      0 (25) 75 & 65.08 (0\%) & 100.28 (75\%) & 81.53 (50\%) \\
      0, 10, 40, 70 & 65.08 (0\%) & 65.08 (0\%) & 69.74 (10\%) \\ 
     \hline
    \end{tabular}
    \caption{River Nidd dataset selected thresholds (and quantile \%) for each method for different grids of candidate thresholds. The Grid column gives \textit{start (increment) end} for each grid.}
    \label{tab: Nidd_grids}
\end{table}

Comparing thresholds selected between the methods is complicated due to the sensitivity of the Wadsworth and Northrop methods to the grid choice. For the EQD, it is natural to use the densest and widest grid, giving $\hat{u}=67.10$. This threshold, which is lower than previously found, gives far more data for the extreme value analysis. As all the River Nidd data are ``extreme", we believe taking $u$ so close to the lower endpoint of the data is not problematic, and it may indicate that the pre-processing level used to produce these data was too high. The first estimated threshold from the Wadsworth (Northrop) methods, without convergence or warning issues,
is $\hat{u}=149.10$ ($\hat{u}=
65.08$). For these three threshold choices, the corresponding GPD parameter estimates (and 95\% CIs) are: $\hat{\sigma}_{u:EQD}= 23.74~(17.78, 29.70)$ and $\hat{\xi}= 0.26~(0.06, 0.46)$ for the EQD; $\hat{\xi}= -0.15~(-1.00, 0.70)$ for the Wadsworth method; and for the Northrop method, 
$\hat{\xi}= 0.20~(0.02, 0.38)$, where we omit the latter two scale parameter estimates as they are estimating different quantities which depend on the threshold, see Section~\ref{section: background}. Provided all estimated thresholds are high enough for the GPD to be appropriate, the values of $\hat{\xi}$ should be similar across methods, due to the threshold stability property (see supplementary material~\ref{section: Sbackground}). The Wadsworth method leads to an extremely wide CI, which results in meaningless inference. However, the EQD and Northrop findings 
about $\xi$
are very similar, but the sensitivity to the candidate grid is still a problem for the Northrop method.

\begin{figure}[ht]
    \centering
    \includegraphics[width=0.8\textwidth]{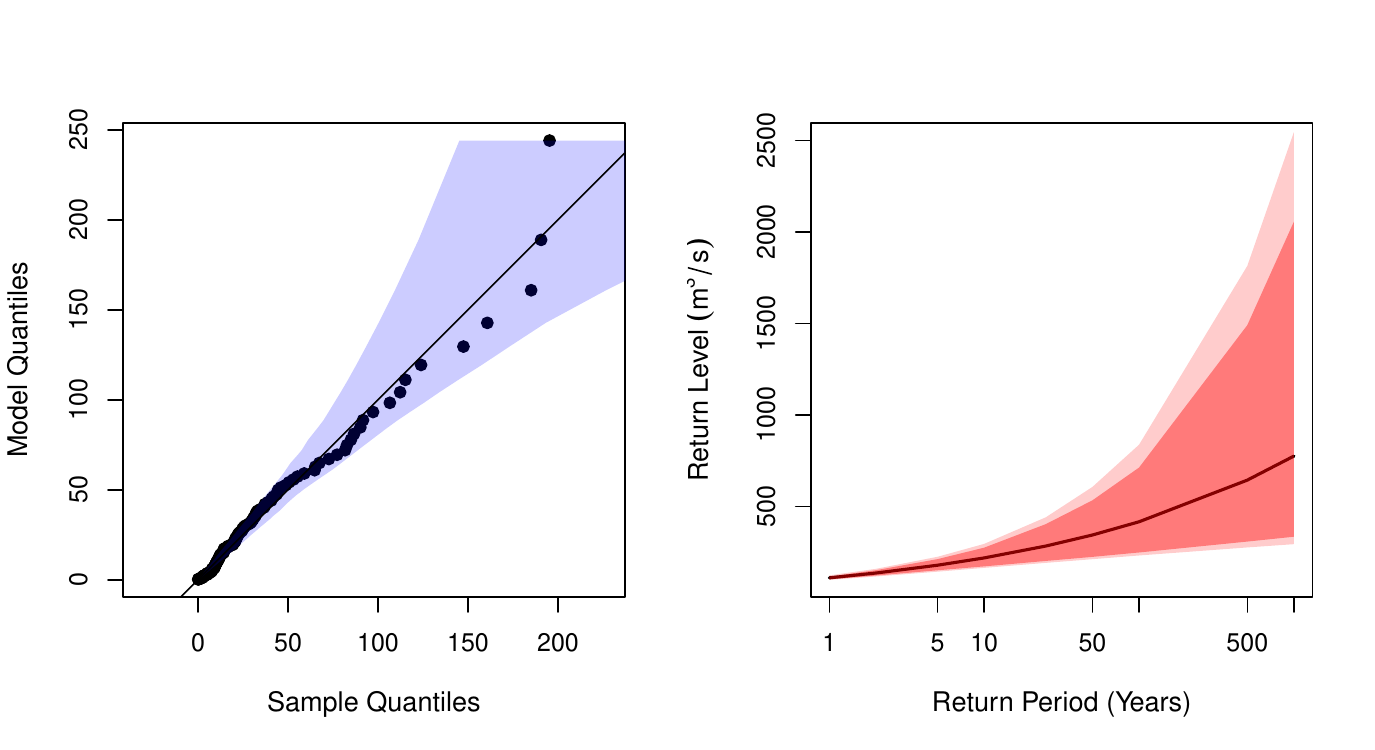}
    \caption{River Nidd analysis: QQ-plot [left] showing model fit with 95\% tolerance bounds [shaded] and return level plot [right] based on EQD threshold choice with 95\% CIs incorporating parameter uncertainty [dark-shaded] and additional threshold uncertainty [light-shaded].}
    \label{fig: return_plot_nidd}
\end{figure}

Figure~\ref{fig: return_plot_nidd} shows a QQ-plot for the GPD model 
using the EQD estimate $\hat{u}=67.10$. The tolerance bounds show a reasonable agreement between model and data. 
For $\hat{u}$, Figure~\ref{fig: return_plot_nidd} also shows the $T$-year return level estimates, with $1\le T \le 1000$. The 95\% CIs incorporate parameter uncertainty alone  and both parameter and threshold uncertainty via Algorithms~1 and 2 respectively, with an increase in uncertainty from the latter for larger $T$; e.g., for the 100- and 1000-year return levels, the CI width increases by a factor of 1.38 and 1.52 respectively. This reiterates how vital it is to incorporate threshold uncertainty into inference.

\section{Conclusion and discussion}
We proposed two substantial developments to 
univariate extreme value analysis.
Firstly, we addressed the widely-studied problem of how to automatically select/estimate a threshold above which an extreme value, generalised Pareto, model can be fitted. We presented a novel and simple approach, which we termed the EQD method, that minimises an approximation to the IAE of the model quantiles and quantiles of the data generating process.
Secondly, we proposed a new  approach to improve the calibration of confidence intervals for high quantile inference, addressing an important but under-studied problem. We achieve this through an intuitively simple, but computationally intensive, double-bootstrapping technique which propagates the uncertainty in the threshold estimation  through to quantile inference.

Regarding the threshold selection component of the work,
we compared the EQD method to the leading existing threshold selection methods in terms of both threshold selection and consequent high quantile estimation. This was conducted using data from iid continuous univariate random variables and the superiority of the EQD method was illustrated across a range of examples using various metrics. Relative to existing approaches, we showed that the EQD exhibits greater robustness to changes in the set of candidate thresholds, to tuning parameters, and avoids a reliance on asymptotic theory in existing likelihood methods. The EQD method is applicable for all data set sizes and for any set of candidate thresholds.

So why does the EQD method perform much better than the existing approaches? Our analysis has identified two core reasons: the choice of a robust measure of goodness of fit for a given (bootstrapped) sample, which controls bias; and the use of bootstrapped replicates, which leads to reduced variance and also appears to reduce bias. Specifically, in comparison to existing methods, the use of our goodness-of-fit measure, over simply exploiting the GPD threshold stability property, ensures better model fits and hence better threshold selection, and the bootstrapping removes the variation that arises if only the observed sample is used, as that  may not be a typical realisation from the underlying data generating process.

In assessing our suggested improvement for the calibration of confidence intervals, we compared the coverage of true quantiles using our proposed approach and the widely-adopted approach of incorporating the GPD parameter uncertainty alone in quantile inference once a threshold has been selected. We showed that the coverage of the existing approach was substantially less than the nominal confidence levels and  our proposed approach led to much more reliable confidence intervals without an undue increase in their width.

While this paper has demonstrated the effectiveness of both the EQD method and our proposed approach for confidence interval construction in the univariate iid setting, we believe that the findings suggest that these approaches could have much wider utility. For example, the \citet{varty2021inference} method, which motivated the structure of the EQD method, was originally developed for non-identically distributed data, with the transformation of excesses of a time-varying threshold to a common marginal Exponential(1) distribution. As such cases typically find that excesses have a common shape parameter $\xi$ \citep{ChavesDavison2005}, we could use the EQD variant of \citet{varty2021inference} by transforming instead to a common GPD with parameters $(1,\xi)$ given we have seen here that by retaining the scale of the original data, the EQD out-performs the \citet{varty2021inference} approach.  
We also believe that the strategy of our new methods could be used to improve threshold estimation in multivariate extremes, in cases of multivariate regular variation assumptions \citep{wan2019threshold} or for asymptotically independent variables \citep{heffernan2004conditional}, and  allow for the uncertainty in this threshold estimation to be incorporated in the subsequent joint tail inferences. Such developments would naturally have similar implications for spatial extreme value modelling as the threshold selection in this context currently comes down to a multivariate (at the data sites) threshold selection process.

\section*{Acknowledgements}
\if0\blind{We are grateful to Ross Towe (Shell) and Peter Atkinson (Lancaster University) for their support and comments.} \fi We also thank the referees and editors for very helpful comments that have improved the presentation and computational evidence of the work.



\bibliographystyle{apalike}
\bibliography{references}

\begin{thebibliography}{}

\bibitem[Bader et~al., 2018]{bader2018automated}
Bader, B., Yan, J., and Zhang, X. (2018).
\newblock Automated threshold selection for extreme value analysis via ordered goodness-of-fit tests with adjustment for false discovery rate.
\newblock {\em Annals of Applied Statistics}, 12(1):310--329.

\bibitem[Belzile et~al., 2023]{belzile2023}
Belzile, L., Dutang, C., Northrop, P., and Opitz, T. (2023).
\newblock A modeler’s guide to extreme value software.
\newblock {\em Extremes}, 26:1--44.

\bibitem[Chavez-Demoulin and Davison, 2005]{ChavesDavison2005}
Chavez-Demoulin, V. and Davison, A.~C. (2005).
\newblock Generalized additive modelling of sample extremes.
\newblock {\em Journal of the Royal Statistical Society: Series C}, 54(1):207--222.

\bibitem[Coles, 2001]{coles2001introduction}
Coles, S.~G. (2001).
\newblock {\em An {I}ntroduction to {S}tatistical {M}odeling of {E}xtreme {V}alues}.
\newblock Springer, New York.

\bibitem[Coles and Pericchi, 2003]{coles2003anticipating}
Coles, S.~G. and Pericchi, L.~R. (2003).
\newblock Anticipating catastrophes through extreme value modelling.
\newblock {\em Journal of the Royal Statistical Society: Series C}, 52(4):405--416.

\bibitem[Coles et~al., 2003]{coles2003fully}
Coles, S.~G., Pericchi, L.~R., and Sisson, S. (2003).
\newblock A fully probabilistic approach to extreme rainfall modeling.
\newblock {\em Journal of Hydrology}, 273(1-4):35--50.

\bibitem[Danielsson et~al., 2001]{danielsson2001using}
Danielsson, J., de~Haan, L., Peng, L., and de~Vries, C.~G. (2001).
\newblock Using a bootstrap method to choose the sample fraction in tail index estimation.
\newblock {\em Journal of Multivariate Analysis}, 76(2):226--248.

\bibitem[Danielsson et~al., 2019]{danielsson2019}
Danielsson, J., Ergun, L., de~Haan, L., and de~Vries, C.~G. (2019).
\newblock {Tail index estimation: quantile-driven threshold selection}.
\newblock Staff Working Papers 19-28, Bank of Canada.

\bibitem[Davison and Smith, 1990]{davison1990models}
Davison, A.~C. and Smith, R.~L. (1990).
\newblock Models for exceedances over high thresholds (with discussion).
\newblock {\em Journal of the Royal Statistical Society: Series B}, 52(3):393--425.

\bibitem[Gelfand, 1996]{gelfand1996model}
Gelfand, A.~E. (1996).
\newblock Model determination using sampling-based methods.
\newblock {\em Markov Chain Monte Carlo in Practice}, 4:145--161.

\bibitem[Gomes, 1994]{gomes1994penultimate}
Gomes, M.~I. (1994).
\newblock Penultimate behaviour of the extremes.
\newblock {\em Extreme Value Theory and Applications: Proceedings of the Conference on Extreme Value Theory and Applications, Gaithersburg Maryland 1993}, 1:403--418.

\bibitem[Heffernan and Tawn, 2004]{heffernan2004conditional}
Heffernan, J.~E. and Tawn, J.~A. (2004).
\newblock A conditional approach for multivariate extreme values (with discussion).
\newblock {\em Journal of the Royal Statistical Society Series B}, 66(3):497--546.

\bibitem[Hill, 1975]{hill1975simple}
Hill, B.~M. (1975).
\newblock A simple general approach to inference about the tail of a distribution.
\newblock {\em The Annals of Statistics}, 3(5):1163--1174.

\bibitem[Murphy et~al., 2023]{Murphy_Software_for_threshold_2023}
Murphy, C., Tawn, J.~A., Varty, Z., and Towe, R. (2023).
\newblock {Software for threshold selection}.
\newblock \url{https://github.com/conor-murphy4/automated_threshold_selection}.

\bibitem[Naveau et~al., 2016]{naveau2016modeling}
Naveau, P., Huser, R., Ribereau, P., and Hannart, A. (2016).
\newblock Modeling jointly low, moderate, and heavy rainfall intensities without a threshold selection.
\newblock {\em Water Resources Research}, 52(4):2753--2769.

\bibitem[Northrop and Attalides, 2020]{threshr}
Northrop, P.~J. and Attalides, N. (2020).
\newblock {\em threshr: Threshold Selection and Uncertainty for Extreme Value Analysis}.
\newblock R package version 1.0.3.

\bibitem[Northrop et~al., 2017]{northrop2017cross}
Northrop, P.~J., Attalides, N., and Jonathan, P. (2017).
\newblock Cross-validatory extreme value threshold selection and uncertainty with application to ocean storm severity.
\newblock {\em Journal of the Royal Statistical Society: Series C}, 66(1):93--120.

\bibitem[Northrop and Coleman, 2014]{northrop2014improved}
Northrop, P.~J. and Coleman, C.~L. (2014).
\newblock Improved threshold diagnostic plots for extreme value analyses.
\newblock {\em Extremes}, 17(2):289--303.

\bibitem[Ossberger, 2020]{tea}
Ossberger, J. (2020).
\newblock {\em tea: Threshold Estimation Approaches}.
\newblock R package version 1.1.

\bibitem[Pickands, 1971]{pickands1971two}
Pickands, J. (1971).
\newblock The two-dimensional {P}oisson process and extremal processes.
\newblock {\em Journal of Applied Probability}, 8(4):745--756.

\bibitem[Pickands, 1975]{pickands1975statistical}
Pickands, J. (1975).
\newblock Statistical inference using extreme order statistics.
\newblock {\em Annals of Statistics}, 3(1):119--131.

\bibitem[Scarrott and MacDonald, 2012]{scarrott2012review}
Scarrott, C. and MacDonald, A. (2012).
\newblock A review of extreme value threshold estimation and uncertainty quantification.
\newblock {\em REVSTAT--Statistical Journal}, 10(1):33--60.

\bibitem[Smith, 1987]{smith1987approximations}
Smith, R.~L. (1987).
\newblock Approximations in extreme value theory.
\newblock Technical report, No. 205, Department of Statistics, Univeristy of North Carolina.

\bibitem[Smith, 2003]{smith2003statistics}
Smith, R.~L. (2003).
\newblock Statistics of extremes, with applications in environment, insurance, and finance.
\newblock In {\em Extreme Values in Finance, Telecommunications, and the Environment}, edited by Finkenstadt, B. and Rootz{\'e}n, H., pages 20--97. Chapman and Hall/CRC.

\bibitem[Varty et~al., 2021]{varty2021inference}
Varty, Z., Tawn, J.~A., Atkinson, P.~M., and Bierman, S. (2021).
\newblock Inference for extreme earthquake magnitudes accounting for a time-varying measurement process.
\newblock {\em arXiv:2102.00884}.

\bibitem[Wadsworth, 2016]{wadsworth2016exploiting}
Wadsworth, J.~L. (2016).
\newblock Exploiting structure of maximum likelihood estimators for extreme value threshold selection.
\newblock {\em Technometrics}, 58(1):116--126.

\bibitem[Wadsworth and Tawn, 2012]{wadsworth2012likelihood}
Wadsworth, J.~L. and Tawn, J.~A. (2012).
\newblock Likelihood-based procedures for threshold diagnostics and uncertainty in extreme value modelling.
\newblock {\em Journal of the Royal Statistical Society: Series B}, 74(3):543--567.

\bibitem[Wan and Davis, 2019]{wan2019threshold}
Wan, P. and Davis, R.~A. (2019).
\newblock Threshold selection for multivariate heavy-tailed data.
\newblock {\em Extremes}, 22(1):131--166.

\end{thebibliography}

\newpage

\setcounter{page}{1}
\setcounter{equation}{0}
\setcounter{section}{0}
\setcounter{figure}{0}
\setcounter{table}{0}

\renewcommand{\theequation}{S.\arabic{equation}}
\renewcommand{\thesection}{S:\arabic{section}}
\renewcommand{\thefigure}{S.\arabic{figure}}
\renewcommand{\thetable}{S.\arabic{table}}




\if0\blind
{
\begin{center}
     {\Large
  \title{\bf Supplementary materials for ``Automated threshold selection and associated inference uncertainty for univariate extremes"\\}}
  \author{Conor Murphy, Jonathan A. Tawn\\
    Department of Mathematics and Statistics, Lancaster University\\
    and \\
     Zak Varty\\
    Department of Mathematics, Imperial College London}
\end{center}
  \maketitle
} \fi

\if1\blind
{
  \bigskip
  \bigskip
  \bigskip
  \begin{center}
    {\LARGE\bf Supplementary Materials for ``Automated threshold selection and associated inference uncertainty for univariate extremes"}
\end{center}
  \medskip
} \fi

\bigskip

\spacingset{2} 

\section{Introduction} 

This document provides further information to accompany the main paper. Section~\ref{section: Sbackground} includes derivations of key properties, further description of methods and additional experiments using some of the standard methods for threshold selection, as referenced in Section~\ref{section: background} of the main paper. Section~\ref{section: Scurrent} provides more detailed descriptions of the \citet{wadsworth2016exploiting} and \citet{northrop2017cross} methods that are used in the analyses of the main paper. 
Section~\ref{section: case_dist} presents further details of the distributions in Cases 1-4 of Section~\ref{section: simstudy} of the main paper, including quantile derivations. Section~\ref{section: supporting_decisions} covers the reasoning for the omission of the \citet{varty2021inference} approach from the simulation study in the main paper, justifies the choice of the default tuning parameters for the EQD method and discusses the choice of calibration data within the definition of $d_b(u)$ in Section~\ref{section: ourmethod} of the main text. Section~\ref{section: extra_sim_res} presents additional simulation experiments, a detailed breakdown of the results outlined in Section~\ref{section: simstudy} of the main paper, and an exploration into the effect of taking different candidate threshold grids. Finally, Section~\ref{section: further_coverage} provides a more extensive analysis of the coverage of true quantiles achieved by both Algorithm 1 and 2, as mentioned in Section~\ref{section: uncertainty} of the main text.

\section{Background for Section~\ref{section: background} material}\label{section: Sbackground}

\subsection{Threshold stability property}\label{section: thresh_stab}

The threshold stability property provides the basis for the standard threshold selection methods described in Section~\ref{section: background} in the main text. This property is derived as follows. Suppose we have that excesses above a threshold $u$ follow a GPD such that $X-u|X>u \sim \text{GPD}(\sigma_u,\xi)$. We want to find the distribution of excesses of a higher threshold $v>u$. Thus, to find the distribution of $X-v|X>v$ for $v$ such that $u<v<x^F$, where $x^F$ is the upper endpoint of the distribution of $X$ and $x>0$, we have:
\begin{equation}
\begin{split}
    \mathbb{P}(X-v > x|X>v) & =  \frac{\mathbb{P}(X>v+x)}{\mathbb{P}(X>v)}
    =  \frac{\mathbb{P}(X>v+x,X>u)}{\mathbb{P}(X>v,X>u)} \\
    & =  \frac{\mathbb{P}(X>v+x|X>u)}{\mathbb{P}(X>v|X>u)} \\
     & = \frac{\left[1 + \frac{\xi (v+x-u)}{{\sigma}_u}\right]_+^{-1/\xi}}{\left[1 + \frac{\xi (v-u)}{{\sigma}_u}\right]_+^{-1/\xi}} \hspace{0.8em} (\text{as} \hspace{0.3em} X-u|X>u \sim \text{GPD}({\sigma}_u,\xi))\\
    & = \left[ 1 + \frac{\xi x}{{\sigma}_u + \xi(v-u)}\right]_+^{-1/\xi}.
\end{split}
\label{eqn:threshStability}
\end{equation}
This is the survivor function of a GPD with scale and shape parameters $({\sigma}_u + \xi(v-u), \xi)$ respectively. Thus, $X-v|X>v \sim \text{GPD}({\sigma}_u+\xi(v-u),\xi)$. 

\subsection{Parameter stability plots}\label{section: parstab}

\begin{figure}[h!]
    \centering
    \includegraphics[width=0.95\textwidth]{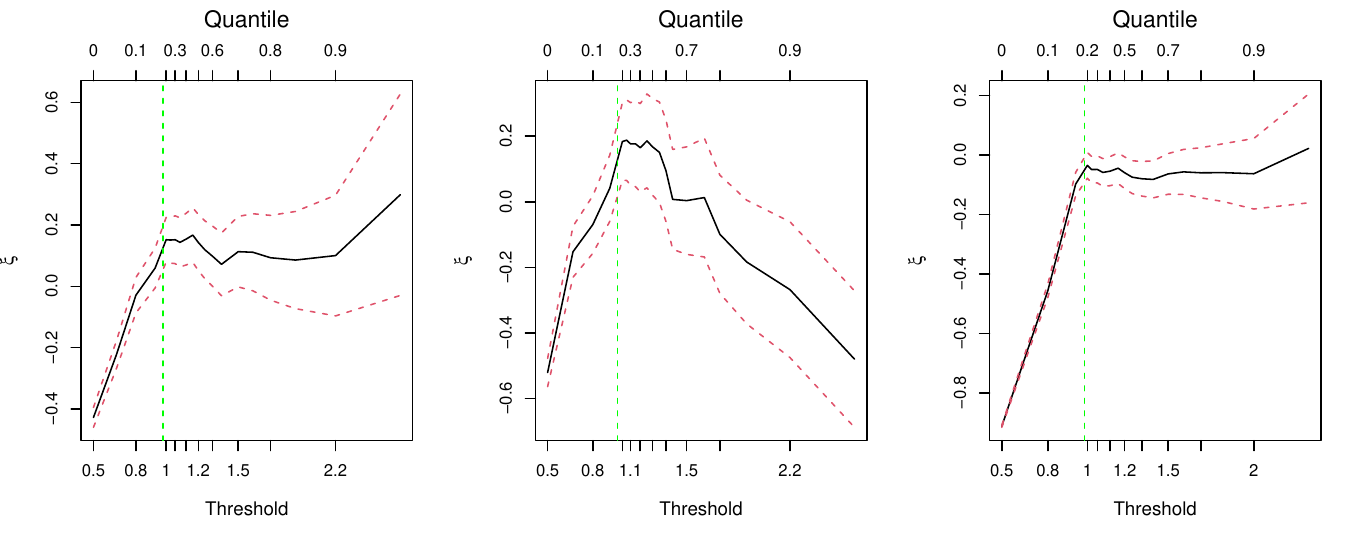}
    \caption{Examples of parameter stability plots based on three simulated datasets used in simulation study in Section~\ref{section: simstudy} of the main text. Solid lines are point estimates (interpolated),  dotted lines are pointwise confidence intervals (interpolated), vertical dashed line is the true threshold.}
    \label{fig: parstab}
\end{figure}

Figure~\ref{fig: parstab} provides three further examples of parameter stability plots to accompany Figure~\ref{fig: parstabexamples} of the main text. The results come from three simulated datasets used in Section~\ref{section: simstudy} of the main paper, specifically the first simulated sample from each of Cases 1-3. These datasets all have true underlying threshold at $u=1.0$, plotted as a vertical green dashed line. The sample sizes of the data analysed in the plots  are $n=1200, 480, 2400$ from left to right. The three datasets were assessed for stability in the shape parameter estimates $\hat{\xi}$ for a grid of candidate threshold choices at sample quantile levels of $0\%, 5\%, \ldots, 95\%$.  The first and third plot seem to show approximate stability above the true threshold. However, the increasing uncertainty as the threshold choice is raised demonstrates the difficulty in objectively assessing this stability. This difficulty is more evident in the centre plot of Figure~\ref{fig: parstab} due to the smaller sample size for these data. The smaller sample size, which is certainly not unusually small for an extreme value analysis, results in highly variable parameter estimates across candidate thresholds and a level of uncertainty which makes the visual assessment of stability difficult. These simulated examples are cases where the underlying true threshold should be fairly clear, and yet, it is not a straightforward task to make appropriate inferences from the parameter stability plots. 

The major criticism of the parameter stability plots is their lack of interpretability, since pointwise confidence intervals are highly dependent across the set of candidate thresholds and, thus, are difficult to account for when assessing stability \citep{wadsworth2012likelihood, northrop2014improved, wadsworth2016exploiting}. The parameter stability plot relies heavily on the chosen grid of candidate thresholds that are compared. Estimates of the shape parameter and confidence intervals are only evaluated at each candidate threshold choice in the grid, meaning that interpretation of a parameter stability plot and the threshold choice itself can be sensitive to the grid of candidate thresholds.  

\subsection{\citet{northrop2014improved}}\label{section: NC_diag}

Here, we provide further description of the \citet{northrop2014improved} method. We describe some of the inadequacies which both limit the applicability of this method and complicate the interpretation of the associated threshold selection plots. 

\begin{figure}[H]
    \centering
    \includegraphics[width=0.95\textwidth]{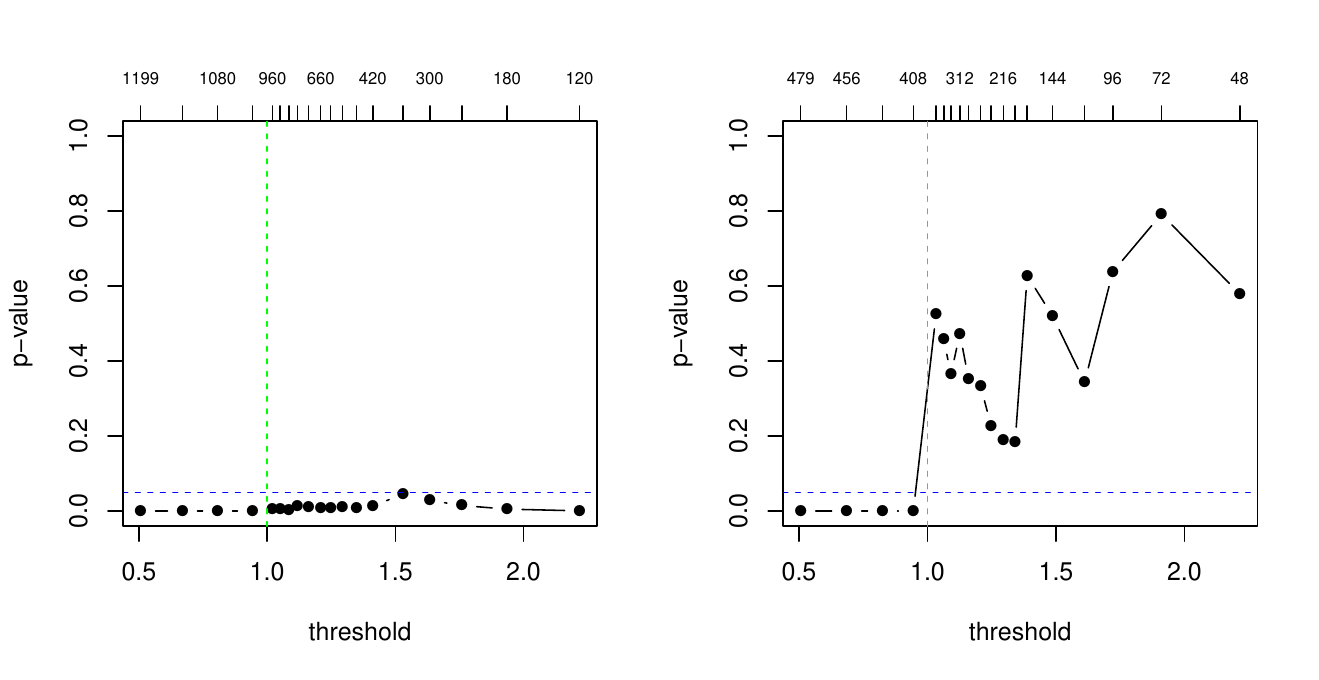}
    \caption{$p$-values derived from \citet{northrop2014improved} method for threshold selection applied to the first simulated samples from Case 1 and 2. Vertical dashed line is the true threshold and the horizontal dashed line shows a $p$-value of $p=0.05$. The numbers above the plot correspond to the numbers of exceedances for each candidate threshold.}
    \label{fig: northropcoleman}
\end{figure}

\citet{northrop2014improved} start by testing the hypothesis that the underlying shape parameter is constant for any threshold above a selected candidate threshold. They extend the piecewise-constant model for the shape parameter of \citet{wadsworth2012likelihood} to allow for an arbitrary number of thresholds and avoid the multiple-testing issue. For their model, they derive likelihood ratio and score methods to test for equality of the shape parameter estimates above a candidate threshold. The method results in a plot of $p$-values against each candidate threshold $u$. An example of the plot of $p$-values obtained using from the \citet{northrop2014improved} method is given in Figure~\ref{fig: northropcoleman}.

Figure~\ref{fig: northropcoleman} shows two plots of the $p$-values derived from the first samples generated in Cases 1 and 2 respectively, described in Section~\ref{section: case_dist}, with candidate thresholds given at the sample $0\%, 5\% \ldots, 95\%$-quantiles (black points) and the true threshold of $u=1.0$ (16.67\% quantile), plotted as a green dashed line. While the \citet{northrop2014improved} method was aimed to improve upon the parameter stability plot in terms of interpretability, it still suffers from problems with subjectivity. For example, whether to select the threshold as the lowest candidate threshold for which the $p$-value rises above the significance level, of say 0.05, or as the candidate threshold which causes the largest increase in the $p$-value.
In the second plot of Figure~\ref{fig: northropcoleman}, both of these approaches would lead us to choose a threshold at the 20\%-quantile which lies near the true value. However, the variability of the $p$-values above casts doubt on this choice. We want to select a threshold where the $p$-values indicate strong evidence for parameter stability for all higher thresholds, excluding very high quantiles where uncertainty may become too large. In Figure~\ref{fig: northropcoleman}, beyond the true threshold, the $p$-values decrease and remain at relatively low levels until another spike at the 65\%-quantile level. Hence, a user of the \citet{northrop2014improved} method may choose a threshold at the 65\%-quantile for this dataset, but even above this value, there is another drop in the $p$-values which may lead the user to an even higher choice, leaving very few exceedances. While the choice is not clear-cut, the outputted $p$-values at least could lead the reader to inspect near the true threshold in this case. 
The same cannot be said for the first plot, where the $p$-values give no indication of a good choice of threshold for this sample. The one value which shows a slight rise in the $p$-value would lead to a threshold choice far from the truth. Whatever the reason for the poor performance in this specific case, while the \citet{northrop2014improved} method tackles some of the inadequacies of parameter stability plots, it suffers from similar shortcomings to the parameter stability plotting method due to the difficulty of interpretation of the resulting plot of $p$-values.

\subsection{Semi-parametric methods}
\label{section: SPDanielsson}
\citet{danielsson2001using,danielsson2019} semi-parametric methods assume that $\xi>0$. The former selects the threshold by minimising the asymptotic mean squared error (MSE) of the Hill estimator \citep{hill1975simple} through a double-bootstrap procedure. The first bootstrap stage computes the optimal size $n_1$ for their second bootstrap stage, where $n_1<n$ and $n$ is the data sample size. To reduce computations, the \textit{tea} package \citep{tea} fixes $n_1=0.9n$. The reliance on asymptotic theory leads to inadequate finite sample performance. The latter method picks the threshold to minimise the \textit{maximum} distance between the empirical and modelled quantiles, i.e., the distance from the diagonal of a QQ-plot. As 
the largest such deviations occur at the highest quantiles and 
the method fails to account for uncertainty, which changes across candidate thresholds, this method over-estimates the threshold.

For the results for the 
\citet{danielsson2001using}
method given in supplementary material \ref{section: detailed_sim_res}, we utilised the \textit{tea} package \citep{tea}, i.e., the package was not built by the authors of the paper; whereas for the \citet{danielsson2019} method, as there did not seem to be code freely available, we constructed our own function.

\section{Further details of existing automated methods}\label{section: Scurrent}

\subsection{\citet{wadsworth2016exploiting}}\label{section: wads2016}
\citet{wadsworth2016exploiting} aims to address the subjective nature of the standard parameter stability plots by utilising the asymptotic distributional theory of the joint distribution of maximum likelihood estimators (MLEs) from samples of exceedances over a range of thresholds. By construction, the exceedances of threshold $v$ are a subset of that of any candidate threshold $u$ whenever $v > u$. Thus, due to this data-overlap, non-standard statistical testing is required, as the data overlap induces dependence between estimates at different thresholds. The method aims to output more interpretable diagnostic plots to improve standard parameter stability plots, primarily by removing dependence between estimates at different candidate thresholds. A simple likelihood-based testing procedure is suggested to allow automated selection of the threshold.

\citet{wadsworth2016exploiting} used the point process representation of extremes, derived in \citet{pickands1971two}, which considers exceedances of a high threshold $u$ as a realisation from a non-homeogeneous Poisson process (NHPP). The representation is outlined as follows. Let $\bm{X} = (X_1,\ldots,X_n)$ be a sequence of independent and identically distributed random variables with common distribution function $F$. Suppose there exists normalising sequences $\{a_n > 0\}$ and $\{b_n \in \mathbb{R}\}$ such that the sequence of point processes
\begin{equation}
\label{eqn: pprep}
    P_n = \left\{\frac{X_i-b_n}{a_n}:i = 1,\ldots,n\right\} \xrightarrow{d} \mathcal{P}
\end{equation}
as $n \rightarrow \infty$ with $\mathcal{P}$ non-degenerate, on the interval $(b_l = \text{lim}_{n \rightarrow \infty} (x_F - b_n)/a_n, \infty)$ where $x_F := \text{inf}\{x : F(x) > 0\}$ is the lower end-point of $F$. Then, $\mathcal{P}$ is a NHPP with intensity 
$\lambda_{\bm{\theta}}(x)$, for $x>b_l$,
and integrated intensity
$\Lambda_{\bm{\theta}}(A)$ on $A=(u,\infty)$, with $u>b_l$, where
\begin{equation*}
    \lambda_{\bm{\theta}}(x) = \frac{1}{\sigma}\left[1 + \xi\left(\frac{x-\mu}{\sigma}\right)\right]_+^{-1-1/\xi}
    \mbox{ and }
\Lambda_{\bm{\theta}}(x) = \left[1 + \xi\left(\frac{x-\mu}{\sigma}\right)\right]_+^{-1/\xi}.
    \end{equation*}
with $\bm{\theta}=(\mu,\sigma,\xi)$, where $\mu \in \mathbb{R}, \sigma > 0, \xi \in \mathbb{R}$ corresponding to the location, scale and shape parameter respectively, with $\xi$ as in the GPD detailed in Section~\ref{section: intro} of the main paper. Hereafter, we let $\bm{\theta_0}$ denote the true value of 
$\bm{\theta}$.

\citet{wadsworth2016exploiting} considers $\bm{x}_N=(x_1,\ldots, x_N)$ as a realisation from a NHPP with a random count $N$ on some region $R=[u_1, \infty)$. 
It is assumed that $\bm{x}_N$ are sorted such that $x_i$ is the $i^{\text{th}}$ largest value, i.e., $x_N < \cdots < x_i < \cdots < x_1$.
A set of candidate threshold choices $(u_1, \ldots, u_k)$ with $b_l\le u_1 < u_2 \cdots < u_k$ which define nested regions  $R_1, R_2, \ldots, R_k$ in $R$ such that $R_j = (u_j, \infty)$ for $j=1,\ldots, k$, so $R_k \subset R_{k-1} \subset \cdots \subset R_1 = R$.  Thus, if $x_1, \ldots, x_{N_j}$ are all the observations which lie in the region $R_j$, then,
there are $N_j$ observations in the region $R_j$. The likelihood of the process over the region $R_j$ is then given up to a constant of proportionality by:
\begin{equation*}
    L_{R_j}(\bm{\theta}) :=\left(\prod_{i=1}^{N_j} \lambda_{\bm{\theta}}(x_i)\right)\exp[-\Lambda_{\bm{\theta}}(R_j)].
\end{equation*}

Now, we denote the MLE of $\bm{\theta}$ based only on the data in region $R_j$ by $\hat{\bm{\theta}}_j$ and the $3 \times 3$ Fisher information matrix for this likelihood as $I_j$ with $I^{-1}_j$ its inverse. \citet{wadsworth2016exploiting} considers a superposition of $m$ replicate Poisson processes as $m \rightarrow \infty$ giving the limit result:
\begin{equation*}
    m^{1/2}\begin{pmatrix}
        \hat{\bm{\theta}}_1 - \bm{\theta}_0,
        \hat{\bm{\theta}}_2 - \bm{\theta}_0,
        \ldots,
        \hat{\bm{\theta}}_k - \bm{\theta}_0
    \end{pmatrix}^T
    \xrightarrow{d}
    N_{3k}(\bm{0}, \Sigma),
\end{equation*}
with $\Sigma$ the $3k \times 3k$ covariance matrix given by $\Sigma = (I^{-1}_{\text{min}(i,j)})_{1 \leq i \leq k, 1 \leq j \leq k}$.

\citet{wadsworth2016exploiting} uses the above result to construct a threshold selection procedure. Isolating the shape parameter $\xi$ in the inference, \citet{wadsworth2016exploiting} denote $m^{-1}\{(I_{j+1}^{-1} - I_j^{-1})_{\xi,\xi}\}$ as the asymptotic variance of the estimated increment $\hat{\xi}_j - \hat{\xi}_{j+1}$ where $\hat{\xi}_j$ is the MLE of the shape parameter on the region $R_j$. As these increments have changing variance with $j$, consider instead the standardised increments $\hat{\bm{\xi}}^* = (\hat{\xi}_1^*, \ldots, \hat{\xi}_{k-1}^*)^T$ given by:
\begin{equation}
    \begin{pmatrix}
        \hat{\xi}^*_1,
        \hat{\xi}^*_2,
        \ldots,
        \hat{\xi}^*_{k-1}
    \end{pmatrix}^T
    := m^{1/2}
    \begin{pmatrix}
        \frac{\hat{\xi}_1 - \hat{\xi}_{2}}{((I_2^{-1} - I_1^{-1})_{\xi,\xi})^{1/2}},
        \frac{\hat{\xi}_2 - \hat{\xi}_{3}}{((I_3^{-1} - I_2^{-1})_{\xi,\xi})^{1/2}},
        \ldots,
        \frac{\hat{\xi}_{k-1} - \hat{\xi}_k}{((I_k^{-1} - I_{k-1}^{-1})_{\xi,\xi})^{1/2}}
    \end{pmatrix}^T,
    \label{eqn: whitenoise}
\end{equation}
which have common unit variances over all components. It follows that if the excesses of $u_1$ follow a GPD then, as $m \rightarrow \infty$,  $\hat{\bm{\xi}}^* \rightarrow \bm{Z}$ where $\bm{Z} \sim N_{k-1}(\bm{0}, \bm{1}_{k-1})$ with $\bm{1}_n$ denoting the $n \times n$-dimensional identity matrix. Given these properties, \citet{wadsworth2016exploiting} term $\hat{\bm{\xi}}^*$ as a white-noise process. 

As a result of the penultimate theory of extremes, described by \citet{smith1987approximations} and \citet{gomes1994penultimate}, \citet{wadsworth2016exploiting} explains that departures from the null assumption of the white-noise process~\eqref{eqn: whitenoise} are a direct consequence of too many values from the body of the data (where the GPD is not appropriate) being included in the estimation. This logic suggests that below the lowest appropriate candidate threshold, say $u_j$, the variables $\hat{\xi}_i^*$, $i=1, \ldots, j-1$ might be better approximated by a $N(\beta, \gamma^2)$, where at least one of $\beta\not= 0$ and $\gamma\not= 1$ holds,  than by a standard normal distribution which is the limit distribution if the threshold $u_j$ was correct. Formally, this gives a changepoint model as:
\begin{equation*}
    \xi_i^* \sim N(\beta, \gamma^2) \quad iid, \quad  i=1, \ldots, j-1, \qquad
    \xi_i^* \sim N(0, 1) \quad iid, \quad i=j, \ldots, k-1,
\end{equation*}
where $j, \beta, \gamma$ are unknown.

\citet{wadsworth2016exploiting} maximise the profile likelihood for $\beta$ and $\gamma$ across $j$ and use a likelihood ratio test to assess if this gives a significantly better fit to $\hat{\bm{\xi}}^*$ than the standard normal distribution. A threshold is automatically selected as the candidate threshold $u_j$ which provides the best fit. If there is no evidence of $\hat{\bm{\xi}}^*$ deviating from white-noise, then the lowest candidate threshold is selected, i.e., $u_1$ in this case.

\subsection{\citet{northrop2017cross}}\label{section: north2017}

Consider $\bm{X}=(X_1, \ldots, X_n)$ where $X_i$ are iid and with associated realisations $\bm{x} = (x_1, \ldots, x_n)$,  where $x_1 < \ldots < x_n$. This contrasts with the notation for the \citet{wadsworth2016exploiting} method, however, we keep this to stay aligned with \citet{northrop2017cross} in our explanation. 

\cite{northrop2017cross} consider $u$ as a training threshold and allow for the threshold exceedance rate, denoted by $\lambda_u = \mathbb{P}(X>u)$, to be incorporated with the GPD parameters $(\sigma_u, \xi)$ into the fit. Thus, $\bm{\theta} = (\lambda_u, \sigma_u, \xi)$ and subsequently, in this section, we refer to the tail model as the GPD$(\lambda_u, \sigma_u, \xi)$. Let $\pi_u(\bm{\theta})$ be a prior density for $\bm{\theta}$. Let $\bm{x}_{(-r)} = \{x_i: 1\le i\le n, i \neq r\}$. The posterior density for $\bm{\theta}$ given data $\bm{x}_{(-r)}$ is denoted $\pi_u(\bm{\theta}|\bm{x}_{(-r)})$ with $\pi_u(\bm{\theta}|\bm{x}_{(-r)}) \propto L(\bm{\theta};\bm{x}_{(-r)}, u)\pi_u(\bm{\theta})$ with the likelihood $L$ assumed to take the form:
\begin{equation}
    \begin{gathered}
         L(\bm{\theta};\bm{x}_{(-r)}, u) = \prod_{i:x_i \in \bm{x}_{(-r)}} f_u(x_i|\bm{\theta}), \\
         f_u(x|\bm{\theta}) = (1 - \lambda_u)^{I(x\leq u)}[\lambda_u h(x - u;\sigma_u, \xi)]^{I(x > u)},
    \end{gathered} \label{eqn: likelinorth}
\end{equation}
where $I(w)$ is an indicator function giving 1 if $w$ is true and 0 otherwise, and $h(x;\sigma_u, \xi) = \sigma_u^{-1}[1+\xi x/\sigma_u]_+^{-(1+1/\xi)}$ is the density of a GPD as in Section~\ref{section: intro} of the main text. In the case of $\lambda_u = 0$, $f_u(x|\bm{\theta}) = I(x \leq u)$. Note that, as defined, $f_u(x|\bm{\theta})$ is not a valid density function as it integrates to $\infty$ and it is discontinuous at $x=u$. The use of the term ``density'' is identified in \citet{northrop2017cross} as an abuse of terminology.

\citet{northrop2017cross} aim to compare a set of candidate values for $u$, denoted $(u_1, \ldots, u_k)$ with $u_1 < \cdots < u_k$, by introducing a fixed validation threshold $v \geq u$ and quantifying the predictive ability of the implied GPD$(\lambda_v, \sigma_v, \xi)$ using each candidate threshold $u_i, i=1,\ldots,k$. They select $v = u_k$. Since $v$ is fixed, the performance of each of the candidate thresholds is compared based on the same validation data. 

To undertake comparisons of fit over different candidate thresholds, a slight extension of the threshold stability property, stated in Section~\ref{section: background} of the main paper, is required, i.e., if a GPD$(\lambda_u, \sigma_u, \xi)$ tail model applies at $u$, this implies a GPD$(\lambda_v, \sigma_v, \xi)$ tail model above $v$ where $\sigma_v = \sigma_u + \xi (v-u)$ and $\lambda_v = \lambda_u[1+\xi(v-u)/\sigma_u]^{-1/\xi} $ assuming that $v$ is such that $1+\xi(v-u)/\sigma_u > 0$. 

For the cross-validation scheme, the data $\bm{x}_{(-r)}$ are the training data with $x_r$ the validation data, and this is repeated for each $r=1, \ldots, n$. To assess the threshold choice performance above $v$, they use leave-one-out cross-validation. The cross-validation predictive density for exceedances of the validation level $v$ under model \eqref{eqn: likelinorth}, using the candidate threshold $u_j$, $j=1, \ldots, k-1$, is then given by:
\begin{equation*}
    f_v(x_r|\bm{x}_{(-r)}, u_j) = \int f_v(x_r|\bm{\theta}) \pi_{u_j}(\bm{\theta}|\bm{x}_{(-r)}) \ \mathrm{d} \bm{\theta} , \quad r = 1, \ldots, n.
\end{equation*}

A Monte Carlo estimator for approximating $f_v(x_r|\bm{x}_{(-r)}, u)$ uses a MCMC generated sample of realisations $\bm{\theta}_j^{(-r)}, j=1, \ldots, m$ (where $m$ is a user choice for the run length of the MCMC after convergence has been deemed to have been achieved) from the posterior distribution $\pi_u(\bm{\theta}|\bm{x}_{(-r)})$ through:
\begin{equation*}
    \hat{f}_v(x_r|\bm{x}_{(-r)}, u) = \frac{1}{m} \sum\limits_{j=1}^m f_v(x_r|\bm{\theta}_j^{(-r)})   , \quad r = 1, \ldots, n.
\end{equation*}
This leads to a measure of predictive ability at $v$ given by:
\begin{equation*}
    \hat{T}_v(u) = \sum\limits_{r=1}^n \log \{\hat{f}_v(x_r|\bm{x}_{(-r)}, u)\},
\end{equation*}
which is evaluated over all candidate thresholds choices of $u_1, \ldots, u_k$. Out of these candidate thresholds, \citet{northrop2017cross} select the one which maximises the measure, $\hat{T}_v$.

To improve the computational efficiency of the estimator $\hat{f}_v(x_r|\bm{x}_{(-r)}, u)$ for $r=1,\ldots,n$, \citet{northrop2017cross} use an importance sampling ratio estimator of \citet{gelfand1996model}. This allows for estimation over $r$ using a single sample from the posterior distribution $\pi_u(\bm{\theta}|\bm{x})$. Specifically, for a single sample $\{\bm{\theta}_j, j=1, \ldots, m\}$ from the posterior $\pi_u(\bm{\theta}|\bm{x})$,
\begin{equation*}
    \hat{f}_v(x_r|\bm{x}_{(-r)}, u) = \frac{\sum\limits_{j=1}^m f_v(x_r|\bm{\theta}_j)q_r(\bm{\theta}_j)}{\sum\limits_{j=1}^m q_r(\bm{\theta}_j)}
    = \frac{\sum\limits_{j=1}^m f_v(x_r|\bm{\theta}_j)/f_u(x_r|\bm{\theta}_j)}{\sum\limits_{j=1}^m 1/f_u(x_r|\bm{\theta}_j)},
\end{equation*}
by taking $q_r(\bm{\theta}) = \pi_u(\bm{\theta}|\bm{x}_{(-r)})/\pi_u(\bm{\theta}|\bm{x}) \propto 1/f_u(x_r|\bm{\theta})$. 

\FloatBarrier

\section{Further details for the distributions of Cases 1-4} \label{section: case_dist}

This section provides full details of the true quantile and density functions used for the simulation study for Cases 1-4 in Section~\ref{section: simstudy} of the main paper.

\subsection{Quantile calculations}

\textbf{Case 1-3}: We simulate $X$ from a mixture of a Uniform$(0.5, 1.0)$ distribution and a GPD$(\sigma_u, \xi)$ distribution above the threshold $u=1.0$. Consequently,
$X$ has distribution function:
\begin{equation}
    F_X(x) = 
    \begin{cases}
        \frac{x-0.5}{3} , & 0.5 \leq x \leq 1 \\
        \frac{1}{6} + \frac{5}{6}\left[H(x-1; \sigma_u, \xi)\right] , & x > 1.
    \end{cases}
    \label{eqn: case1-3cdf}
\end{equation}
where $H(x-1; \sigma_u, \xi)$ is the distribution function of a GPD with parameters $(\sigma_u, \xi)$.
Therefore, the true quantile $x_p$ with exceedance probability $p$ (for $p < 5/6$) is
\begin{equation*}
    x_p = 1 + \frac{\sigma_u}{\xi}\left[\left(\frac{6p}{5}\right)^{-\xi} - 1\right].
\end{equation*}
This formulation is identical across Cases 1-3, but the model parameters and simulation sample sizes differ over these cases, with these values being given in Table~\ref{tab: case_descriptions} of Section~\ref{section: simstudy} in the main paper. 

\textbf{Case 4}: 
Here $X$ has distribution function,
for $x>0$, of 
\begin{equation}
     F_X(x) = 
        \int_0^x \frac{h(s; \sigma, \xi) \mathbb{P}(\mathcal B < s)}{q + \bar{H}(1; \sigma, \xi)} \ \mathrm{d}s
    \label{eqn: case2cdf}
\end{equation}
where $\bar{H}(x; \sigma, \xi)$ and $h(x; \sigma, \xi)$ are the survivor and density functions of a GPD with parameters $(\sigma, \xi)$ and a threshold $0$, $\mathcal B \sim \text{Beta}(\alpha, \beta)$, so $0\le \mathcal B\le 1$ and $q=\int_0^1 h(s; 0.5, 0.1) \mathbb{P}(\mathcal B < s) \mathrm{d}s$. 
This unusual distribution function transitions from a non-GPD distribution to an exact GPD for excesses of $1$ (the upper bound of $\mathcal B$), so that the true threshold for $X$ is $u=1$ and the excess distribution, for $x>1$, has survivor function $\bar{H}(x - 1; \sigma_1=\sigma+\xi, \xi)$ following from the threshold stability property~\eqref{eqn:threshStability}.
Here $\tau:=\mathbb{P}(X\leq 1)=q/ (q + \bar{H}(1; \sigma, \xi))$ so the true quantile $x_p$ for $X$ with exceedance probability $p$, where $p\leq 1-\tau$, is:
\begin{equation*}
    x_p = 1 + \frac{\sigma_1}{\xi}\left[\left(\frac{p}{1-\tau}\right)^{-\xi} - 1\right].
\end{equation*}

Although $F_X$ appears complex, it is straightforward to simulate samples from $X$. Specifically, $X$ is generated using a rejection method by first simulating a proposal variable $Y \sim \text{GPD}(\sigma, \xi)$ above the threshold of $0$. 
The rejection step involves rejecting $Y$ as a candidate for $X$ only if $Y<\mathcal B$, where $\mathcal B$ is a random variate generated from a $\text{Beta}(\alpha, \beta)$ distribution. Since $\mathcal B\leq 1$, for $Y>1$ it follows that $X=Y$, but only a proportion of the values of $Y<1$ are retained in $X$, with the rate of retention dependent on the parameters of the $\text{Beta}(\alpha, \beta)$ distribution. 

In our simulations, we selected $(\sigma,\xi)=(0.5,0.1)$ and $(\alpha,\beta)=(1,2)$, with the latter chosen such that the density of $X$ has a mode below $1$. For these parameters
we obtain $\hat{\tau} = 0.721$, to 3 decimal places, where we evaluated $\tau$ using Monte Carlo integration methods. To ensure that we have the same numbers of exceedances of the true threshold across samples, we simulate until we have a sample proportion of threshold exceedances matching the true value of $1 - \tau$ and an overall sample size of 1000.

\subsection{Density and quantile functions}

\begin{figure}[h!]
    \centering
    \includegraphics[width=0.47\textwidth, page= 1]{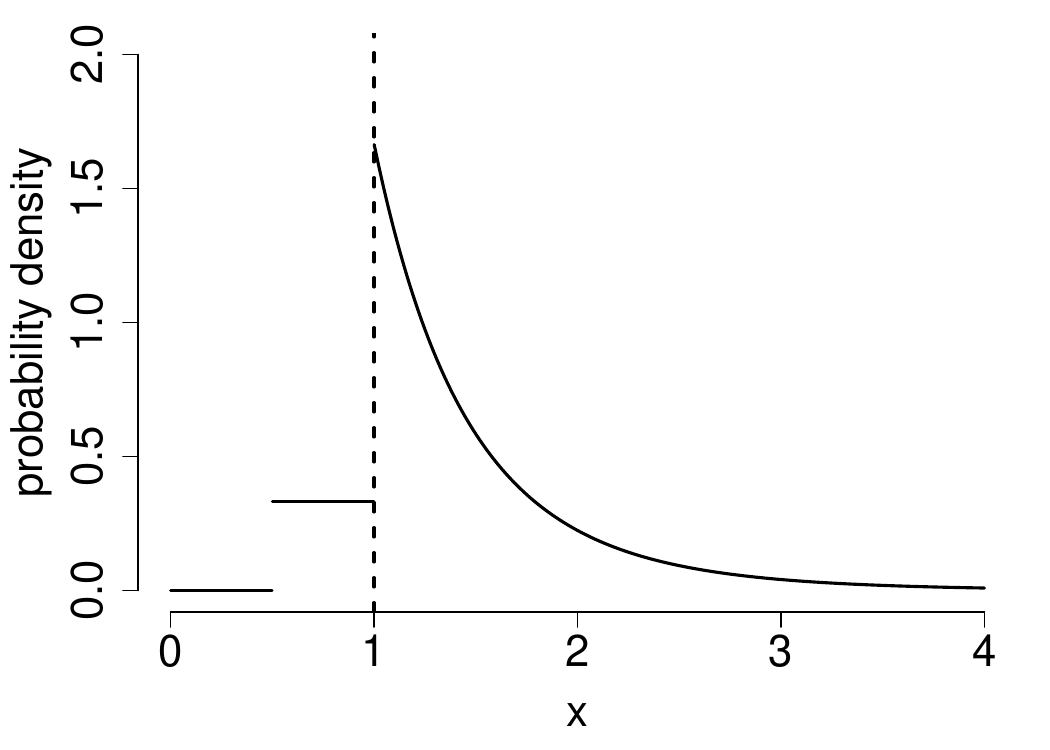} \quad 
    \includegraphics[width=0.47\textwidth, page= 2]{SM_figures/case_study_denisty_plots.pdf} \\
    \includegraphics[width=0.47\textwidth, page= 3]{SM_figures/case_study_denisty_plots.pdf} \quad
    \includegraphics[width=0.47\textwidth, page= 4]{SM_figures/case_study_denisty_plots.pdf}
    \caption{True densities of simulated datasets from Cases 1-4 with numbering corresponding to left-right and then top to bottom.}
    \label{fig: sim_data_ex}
\end{figure}

The density functions  of the random variable $X$ are given in Figure~\ref{fig: sim_data_ex}, for Cases 1-4. 

All four density plots show that there is a large probability of exceeding the true threshold
(i.e., $X>1$) so a large proportion of each sample is from a GPD tail. This is unusual in practice (where often thresholds correspond to 90-99\% sample quantiles) and so threshold selection should be easier in these examples than typically. This is especially true for Cases 1-3 with the density having a large step change at the threshold and the density having a completely different shape below the threshold. 
In contrast, Case 4 has a much more subtle transition for the density across the threshold. The density is continuous and first-differentiable at the threshold, but with a clearly non-GPD distribution below the threshold, shown by the density's mode lying away from its lower endpoint. Thus, we see this case as much more challenging for threshold selection.

\begin{figure}[h]
    \centering
    \includegraphics[width=0.95\textwidth]{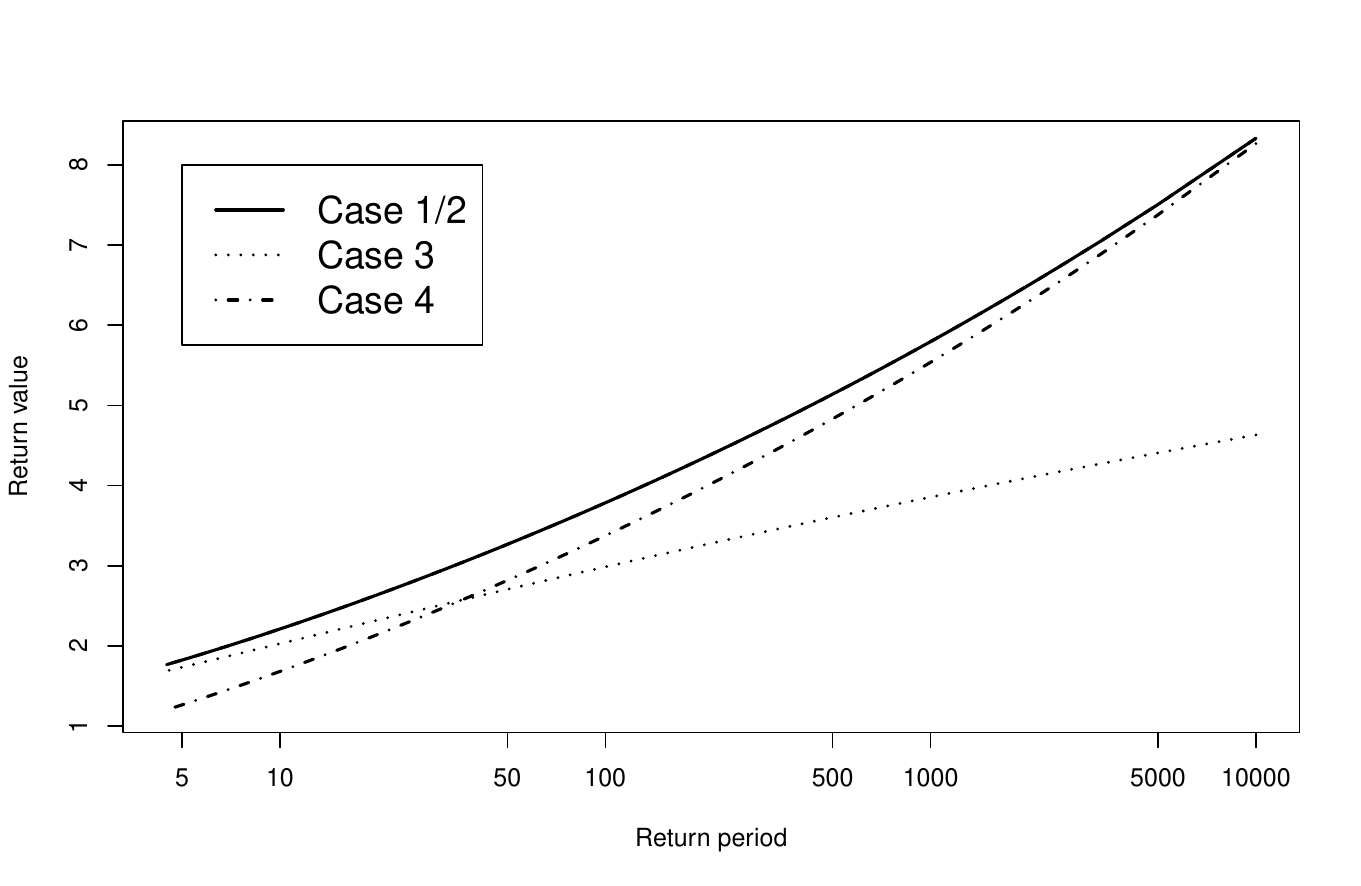}
    \caption{True return values of simulated datasets from Cases 1-4.}
    \label{fig: return_case1-4}
\end{figure}

To further emphasise the differences between Cases 1-4, Figure~\ref{fig: return_case1-4} shows return level plots for the simulated distributions of Cases 1-4 for a range of return periods (per observation). In particular, this plot emphasises the key difference between Case 3 and the other cases; 
Case 3 has a finite upper-endpoint due to $\xi < 0$ whereas the other cases have
unbounded distributions. This difference is not apparent from the density plots 
in Figure~\ref{fig: sim_data_ex}.

The idea behind our choice of distributions is that if methods struggle in cases where we have a clear true threshold then there will be significant problems when it comes to real datasets. So, collectively the four cases provide a natural testing ground for separating between threshold methods.

\section{Supporting details for Section~\ref{section: ourmethod}}\label{section: supporting_decisions}

\subsection{Overview}
In this section, we provide evidence, based on a range of simulation studies, to support several decisions we made in Section~\ref{section: ourmethod} of the main paper. Specifically, in Section~\ref{section: comp_varty}, we provide evidence to demonstrate the advantages of using the EQD over the \citet{varty2021inference} method; Section~\ref{section: sensitivity_analysis} outlines results indicating that our suggested default choices for the tuning parameters $(B,m)$ of the EQD method are widely suitable, and that there is very little sensitivity to these choices; in Section~\ref{subsec:calibrationdata}, we show that the choice of bootstrap data as the calibration data in the metric $d_b(u)$ works comparably relative to using the observed data; and in Section~\ref{section: bootstrapping}, we demonstrate the benefits of the bootstrapping component of the EQD method.

\subsection{Comparison of EQD and \citet{varty2021inference} methods}
\label{section: comp_varty}
This section provides the full evidence basis for the decision, outlined in Section~\ref{subsec:metric}, to prefer the EQD method and omit the results of the \citet{varty2021inference} method from the main paper. Here, we compare both methods using the data simulated from Cases 1-4, as described in detail in Section~\ref{section: case_dist}, and outlined overview in Section~\ref{section: simstudy} of the main paper. As seen from Section~\ref{subsec:metric} of the main paper, the EQD and the \citet{varty2021inference} methods differ simply in the scale on which the metric of goodness-of-fit is compared, with the former evaluated on the observed data scale and the latter making the same comparison after 
transformation onto Exponential(1) margins.

We assess the performance of the two methods based on both the selection of thresholds (as the truth is known in each case) and on the subsequent estimation of quantiles for a range of exceedance probabilities, namely the $(1-p_{j,n})$-quantiles where $p_{j,n} = 1/(10^jn)$ for $j=0,1,2$ with $n$ denoting the length of the simulated dataset. For each case and each of the measures of fit, all comparisons between the methods are based on estimates obtained using the same set of 500 replicated samples, so any difference in the methods found is simply due to the two method's performance.

Table~\ref{tab: EQD_Varty_Case1-4} shows the RMSE, bias and variance of the thresholds chosen by the two methods. Based solely on threshold choice, it is difficult to distinguish between the approaches with each method narrowly outperforming the other in two of the four cases based on each of RMSE and bias. Cases 1-3 exhibit positive bias for both methods, as they are much less likely to pick a threshold too low in these cases given the sudden change in the density shown in Figure~\ref{fig: sim_data_ex}. In contrast, for Case 4, both methods incur a negative bias, essentially due to the smooth transition from GPD in the density shown in Figure~\ref{fig: sim_data_ex}. Additionally, in every case, there is no method that gives threshold estimates of lower variance than the EQD method.

The primary goal of an extreme value analysis is usually quantile inference, rather than threshold selection. Tables~\ref{tab: rmse_quant_varty} and
\ref{tab: bias_var_quant_varty} compare the EQD and \citet{varty2021inference} methods in terms of quantile inference. These tables show, for the three target quantiles, the RMSE, bias and variance of quantile estimates that are based on MLE fits of a GPD above the thresholds selected using each method. 
The differences between the two methods are evident. The EQD either matches or achieves the lower RMSE in all cases and all quantiles and the differential in performance becomes more evident for long-range extrapolation, i.e., as $j$ increases, see Table~\ref{tab: rmse_quant_varty}.
This difference in RMSE seems to stem mainly from the smaller variance of the EQD estimates, see Table~\ref{tab: bias_var_quant_varty} [right].
As with the results for threshold selection, the bias results for quantile estimation in 
Table~\ref{tab: bias_var_quant_varty} [left] show the two methods perform similarly (with each method slightly better on a number of occasions across the cases and the three quantile levels of interest).

Given that the goal of a threshold selection method is to improve high quantile inference, we conclude from this study that, while results are similar, the EQD appears to perform better for quantile estimation. Furthermore, the \citet{varty2021inference} method requires an additional, and non-intuitive (for iid data), transformation to Exponential(1) margins. 
Thus, we choose to omit the \citet{varty2021inference} approach from subsequent studies in the supplementary material and from the simulation study discussed in Section~\ref{section: simstudy} of the main text.

\begin{table}[h!]
    \centering
    \begin{tabular}{|c|c|c|c|c|c|c|}
        \hline
        & \multicolumn{3}{c|}{\textit{EQD}} & \multicolumn{3}{c|}{\textit{Varty method}} \\
        \hline
        Case & RMSE & Bias & Variance & RMSE & Bias & Variance\\
        \hline
        Case 1 & $\bm{0.048}$ & $\bm{0.034}$ & $\bm{0.001}$ & 0.059 & 0.041 & 0.002 \\
        Case 2 & $\bm{0.060}$ & $\bm{0.031}$ & $\bm{0.003}$ & 0.073 & 0.039 & 0.004 \\
        Case 3 & 0.060 & 0.042 & 0.002 & $\bm{0.055}$ & $\bm{0.039}$ & 0.002 \\
        Case 4 & 0.526 & $-0.515$ & $\bm{0.012}$ & $\bm{0.508}$ & $\bm{-0.492}$ & 0.016 \\        \hline
    \end{tabular}
    \caption{Measures of performance (RMSE, bias and variance) for threshold choices  for the EQD and \citet{varty2021inference} methods,
    for Cases 1-4. The smallest magnitude for each measure of performance are highlighted in bold for each case.}
    \label{tab: EQD_Varty_Case1-4}
\end{table}

\begin{table}[h!]
    \centering
    \begin{tabular}{ |c|c|c|c|c| } 
    \hline 
     & \textit{EQD}  & \textit{Varty} &  \textit{EQD} & \textit{Varty} \\
     \hline 
     $j$ & \multicolumn{2}{c|}{\textbf{Case 1}} & \multicolumn{2}{c|}{\textbf{Case 2}} \\
     \hline
     0 & $\bm{0.563}$ & 0.605  &   $\bm{0.599}$ & 0.611  \\ 
     1 & $\bm{1.258}$ & 1.335 & $\bm{1.488}$ & 1.542  \\ 
     2 & $\bm{2.447}$ & 2.612 &  $\bm{3.119}$ & 3.305 \\       
     \hline
    & \multicolumn{2}{c|}{\textbf{Case 3}} & \multicolumn{2}{c|}{\textbf{Case 4}} \\
     \hline
     0 & 0.190 & 0.190 & $\bm{0.677}$ & 0.705  \\ 
     1 & $\bm{0.323}$ & 0.324 & $\bm{1.563}$ & 1.673 \\ 
     2 & 0.483 & 0.483  & $\bm{3.043}$ & 3.378 \\       
     \hline
    \end{tabular}
\caption{RMSE of the estimated $(1-p_{j,n})$-quantiles in Cases 1-4 based on fitted GPD above chosen threshold for the EQD and \citet{varty2021inference} methods. The smallest value for each quantile are highlighted in bold.}
 \label{tab: rmse_quant_varty}
\end{table}

\begin{table}[h!]
    \centering
    \begin{tabular}{ |c|c|c|c|c| } 
    \hline 
     & \textit{EQD}  & \textit{Varty} &  \textit{EQD} & \textit{Varty} \\
     \hline 
     $j$ & \multicolumn{2}{c|}{\textbf{Case 1}} & \multicolumn{2}{c|}{\textbf{Case 2}} \\
     \hline
     0 & $-0.021$ & $\bm{-0.005}$  &  $-0.049$ & $\bm{-0.018}$  \\ 
     1 & $\bm{-0.015}$ & 0.030 & $\bm{-0.046}$ & 0.055  \\ 
     2 & $\bm{0.044}$ & 0.145 &  $\bm{0.069}$ & 0.312 \\       
     \hline
    & \multicolumn{2}{c|}{\textbf{Case 3}} & \multicolumn{2}{c|}{\textbf{Case 4}} \\
     \hline
     0 & $-0.008$ & $\bm{-0.007}$ & $-0.283$ & $\bm{-0.233}$ \\ 
     1 & $-0.007$ & $\bm{-0.005}$ &  $-0.722$ & $\bm{-0.571}$ \\ 
     2 & $-0.002$ & 0.002 &  $-1.410$ & $\bm{-1.064}$ \\       
     \hline
    \end{tabular}
    \quad
        \begin{tabular}{ |c|c|c|c|c| } 
    \hline 
     & \textit{EQD}  & \textit{Varty} &  \textit{EQD} & \textit{Varty} \\
     \hline 
     $j$ & \multicolumn{2}{c|}{\textbf{Case 1}} & \multicolumn{2}{c|}{\textbf{Case 2}} \\
     \hline
     0 & $\bm{0.316}$ & 0.335  &   $\bm{0.357}$ & 0.373  \\ 
     1 & $\bm{1.582}$ & 1.716 & $\bm{2.211}$ & 2.376  \\ 
     2 & $\bm{5.988}$ & 6.638 &  $\bm{9.723}$ & 10.824 \\       
     \hline
    & \multicolumn{2}{c|}{\textbf{Case 3}} & \multicolumn{2}{c|}{\textbf{Case 4}} \\
     \hline
     0 & 0.036 & 0.036 & $\bm{0.379}$ & 0.444  \\ 
     1 & 0.105 & 0.105 & $\bm{1.926}$ & 2.479 \\ 
     2 & 0.233 & 0.233  &  $\bm{7.287}$ & 10.299 \\       
     \hline
    \end{tabular}
\caption{Bias [left] and variance [right] of the estimated 
$(1-p_{j,n})$-quantiles in Cases 1-4 based on fitted GPD above chosen threshold for the EQD and \citet{varty2021inference} methods. The smallest variance and absolute bias for each quantile are highlighted in bold.}
 \label{tab: bias_var_quant_varty}
\end{table}

\newpage
\subsection{Selection of default tuning parameters}
\label{section: sensitivity_analysis}

This section provides the sources of evidence presented in the main paper for Sections~\ref{section:whyEQDchoice} and \ref{subsec:tuning} in terms of the effect of $m$ on the interpolation of quantiles in the metric and the suitability of our default values of the tuning parameters $(B,m)$.

First, consider an analysis of the sensitivity of the EQD method to different choices of its two tuning parameters $(B,m)$, where $B$ denotes the number of the bootstraps for which $d_b(u)$ is evaluated in order to calculate the overall metric $d_E(u)$, and $m$ is the number of quantiles used in the evaluation of the metric $d_b(u)$ for each bootstrap. In the main paper, the values of $(B,m)=(100,500)$ are proposed as the default values for the simulation studies of the performance of the EQD method. These values are used for the tuning parameters in all simulation studies of the EQD method in the main paper and the supplementary material.

We focus our sensitivity analysis on the 500 replicated samples of Case 1 where $n=1000$, detailed in Section~\ref{section: simstudy} of the main paper. Tables~\ref{tab: RMSE_varying_B} \& \ref{tab: RMSE_varying_m} provide the RMSEs of the threshold estimates along with the computation time relative to that of the default value when using the EQD with different values of $B$ and $m$ respectively. 

For the choice of $B$, the number of the bootstraps for which $d_b(u)$ is evaluated when calculating $d_E(u)$, Table~\ref{tab: RMSE_varying_B} shows that the computation time of the EQD method increases linearly with $B$. As $B$ is simply the number of bootstrap samples in an average, then in principle we want to take $B$ as large as possible to remove Monte Carlo noise in the average approximation of the associated expectation. Thus, in selecting $B$, we require it to be sufficiently large so that any residual Monte Carlo noise is not important (for threshold selection to be stable) whilst recognising the linear increase in computation time from this choice. Thus, for one-off analyses, as computation time is not of particular concern, it is ideal to take $B$ as large as possible. However, for simulation studies, a more careful choice of $B$ is required as accuracy needs to be balanced with computation time. Table~\ref{tab: RMSE_varying_B} provides evidence on how the Monte Carlo noise is diminishing as $B$ increases, and shows the RMSE value stabilising as $B$ increases. The decreases in RMSE are only very slight, especially when compared to the differences we see between the EQD and other existing approaches in Section~\ref{section: simstudy} of the main paper. 

\begin{table}[h!]
    \centering
    \begin{tabular}{|c|c|c|c||c|} 
    \hline
     $B$ & 200 & 400 & 1000 & 100 (default) \\
     \hline
      RMSE & 0.043 & 0.039 &  0.040 & 0.048\\ 
      Relative time & 2 & 4 & 10 & 1  \\ 
     \hline
    \end{tabular}
\caption{RMSEs for threshold estimates and for the relative computation time compared to the default choice of $B=100$ obtained using the EQD method for different values of $B$ for Case 1. Each result uses $m=500$ and 
500 replicated samples.}
 \label{tab: RMSE_varying_B}
\end{table}

For the choice of $m$, the number of quantiles used in the evaluation of the metric $d_b(u)$ for each bootstrap, Table~\ref{tab: RMSE_varying_m} shows results for two different strategies for selecting $m$: the first allowing $m$ to be proportional to the data sample size $n$ irrespective of threshold value, i.e. $m=cn$ for $c=0.5,1,2,10$, and the second allowing $m$ to vary according to the number, $n_u$, of exceedances of each candidate threshold $u$, i.e. $m=cn_u$ for $c=0.5,1,2,10$. 
Our reason for exploring the second strategy is to ensure that across candidate thresholds we are using the same level of interpolation/extrapolation to non-sample quantiles. For each strategy we examine the effect of different degrees of proportionality $c$. The RMSE of the threshold estimates obtained using the EQD show little sensitivity to the value of $m$ across the two strategies and all levels of proportionality. 

\begin{table}[h!]
    \centering
    \begin{tabular}{|c|c|c|c|c||c|c|c|c||c|}
    \hline
    & \multicolumn{4}{c||}{$m=cn$} & \multicolumn{4}{c||}{$m=cn_u$} & $m=500$ (default)\\
    \hline
     $c$ & 0.5 & 1 & 2 & 10 & 0.5 & 1 & 2 & 10 & \\
     \hline
      RMSE & 0.045 & 0.046 &  0.045 & 0.046 & 0.049 &  0.048 & 0.047 & 0.047 & 0.048 \\ 
      Relative time & 1.1 & 1.4 & 2.0 & 7.2 & 0.9 & 1.1 & 1.4 & 4.6 & 1\\ 
     \hline
    \end{tabular}
\caption{RMSEs for threshold estimates and for the relative computation time compared to the default of $m=500$ of the EQD method for different values of $m$ for Case 1. Each result uses $B=100$ and 
500 replicated samples.}
 \label{tab: RMSE_varying_m}
\end{table}

Table~\ref{tab: RMSE_varying_m} also shows that increasing $m$ in either strategy is essentially wasting the increased computation time. When estimating $d_b(u)$ for a particular bootstrap, we are aiming to approximate the integrated absolute error (IAE) between the model quantiles and sample quantiles for that sample. This $d_b(u)$ then feeds into the overall $d_E(u)$ for the excesses of candidate threshold $u$ which approximates the IAE between model quantiles and the data generating process. This sensitivity analysis shows that once we choose a suitably large value for $m$, any changes in this approximation error are quite small in comparison to the $d_b(u)$ value itself.

Now consider the effect of $m$ on the interpolation of quantiles in the metric. While we have shown that changing $m$ does not meaningfully affect the RMSE of threshold choice over repeated samples, it is still important to investigate the effect of this choice for the values of $d_b(u)$ and $d_E(u)$ for a range of candidate thresholds as well as the effect, if any, on the resulting threshold choice for a particular dataset. In particular, we are interested in the effect of the choice of interpolation grid between $m=500$ and $m=n_u$. 

For a particular bootstrap sample, the choice of $m=500$ can lead to under- or over-sampling of the upper tail when approximating the IAE of the QQ-plot, depending on if $m<n_u$ or $m > n_u$. While this may not be ideal, it is only important if it has a significant effect on the overall metric value in a way that unfairly or adversely affects the resulting choice of thresholds. 

To explore the effect of the interpolation grid on the sampling distribution of metric values for thresholds (i.e., $d_b(u)$ values), $d_E(u)$ and on the resulting threshold choice, we have considered the following additional investigations. Let $d^{500}_b(u)$ and $d^{n_u}_b(u)$ denote the value of the metric for the $b^{\text{th}}$ bootstrap sample above threshold $u$ using $m=500$ and $m=n_u$ respectively. For the first simulated sample of Case 1 and the Gaussian case, we look at:
\begin{enumerate}
    \item The distribution of $d^{500}_b(u)$ and $d^{n_u}_b(u)$ for each $u$ over the candidate grid of thresholds.
    \item The distribution of the relative difference between $d^{500}_b(u)$ and $d^{n_u}_b(u)$ to the overall metric $d_E(u)$ over the candidate grid.
\end{enumerate}
The reasons that we selected these two features to investigate are that the former looks at the effect of the interpolation on the sampling distribution of the metric while the latter assesses if the interpolation method could significantly change the overall metric value for any particular thresholds on a particular dataset. 

\begin{figure}[h]
    \centering
    \includegraphics[width=0.95\textwidth]{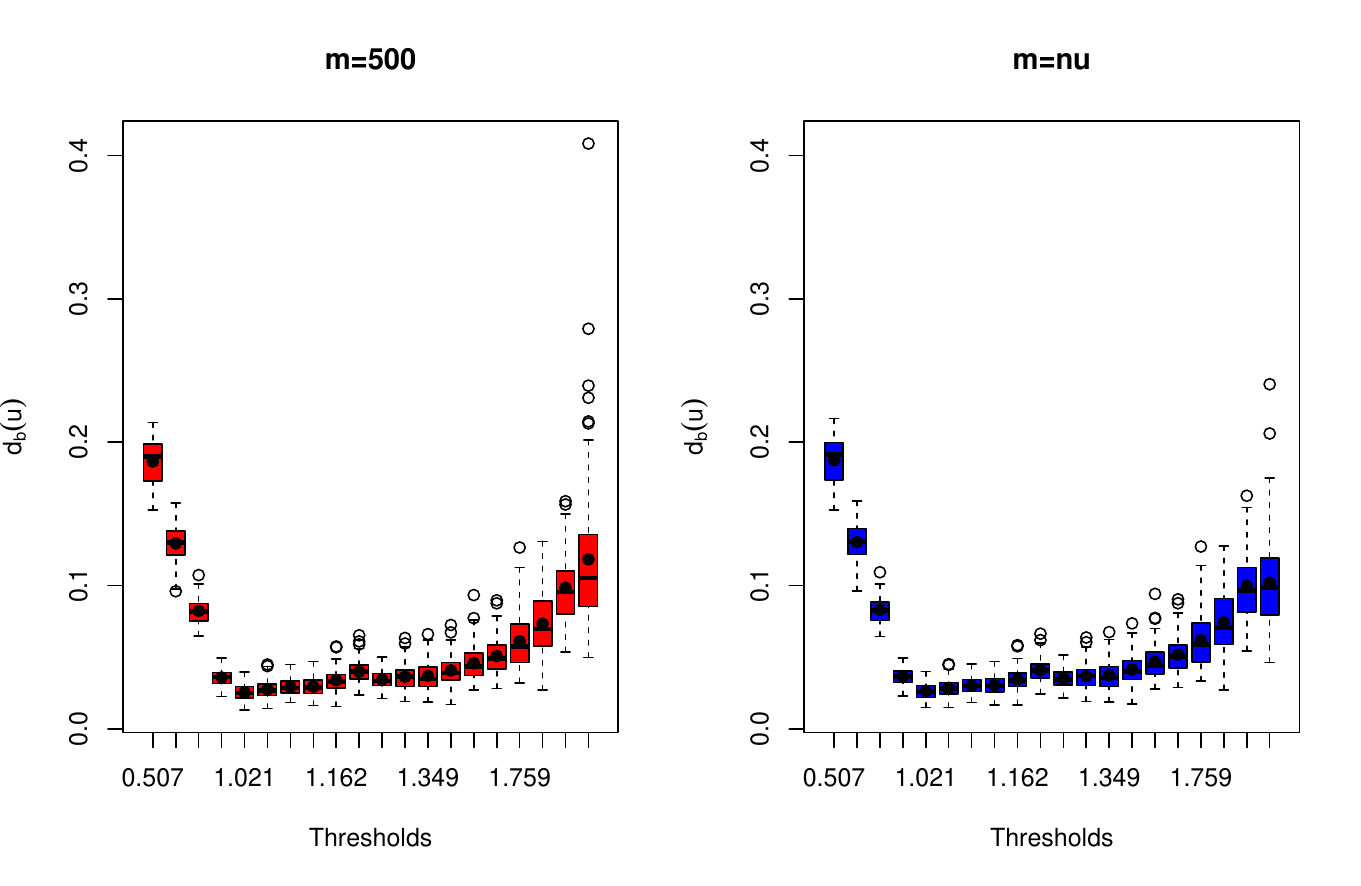}
    \caption{Boxplots of $d^{500}_b(u)$ [left] and $d^{n_u}_b(u)$ [right] for each $u$ over the candidate grid of thresholds for the first sample of Case 1. The mean for each threshold is shown as a black point.}
    \label{fig: m_500_vs_mnu}
\end{figure}

For the first simulated data sample of Case 1, we explore
the distribution of $d^{500}_b(u)$ and $d^{n_u}_b(u)$
in Figure~\ref{fig: m_500_vs_mnu}. Specifically, this figure shows boxplots of the distribution of $d^{500}_b(u)$ 
and $d^{n_u}_b(u)$ 
values for each value of $u$. The mean of these values, i.e.,  $d_E(u)$, 
is also plotted as a black point in each boxplot. For particular thresholds, comparison of the sampling distributions of $d^{500}_b(u)$ and $d^{n_u}_b(u)$ values shows only very minor differences. While there are some larger differences between the plots, particularly at higher thresholds, the black points in each plot indicate that these differences in $d_b(u)$ do not lead to large differences in the overall metric value $d_E(u)$. 

\begin{figure}[H]
    \centering
    \includegraphics[width=0.95\textwidth]{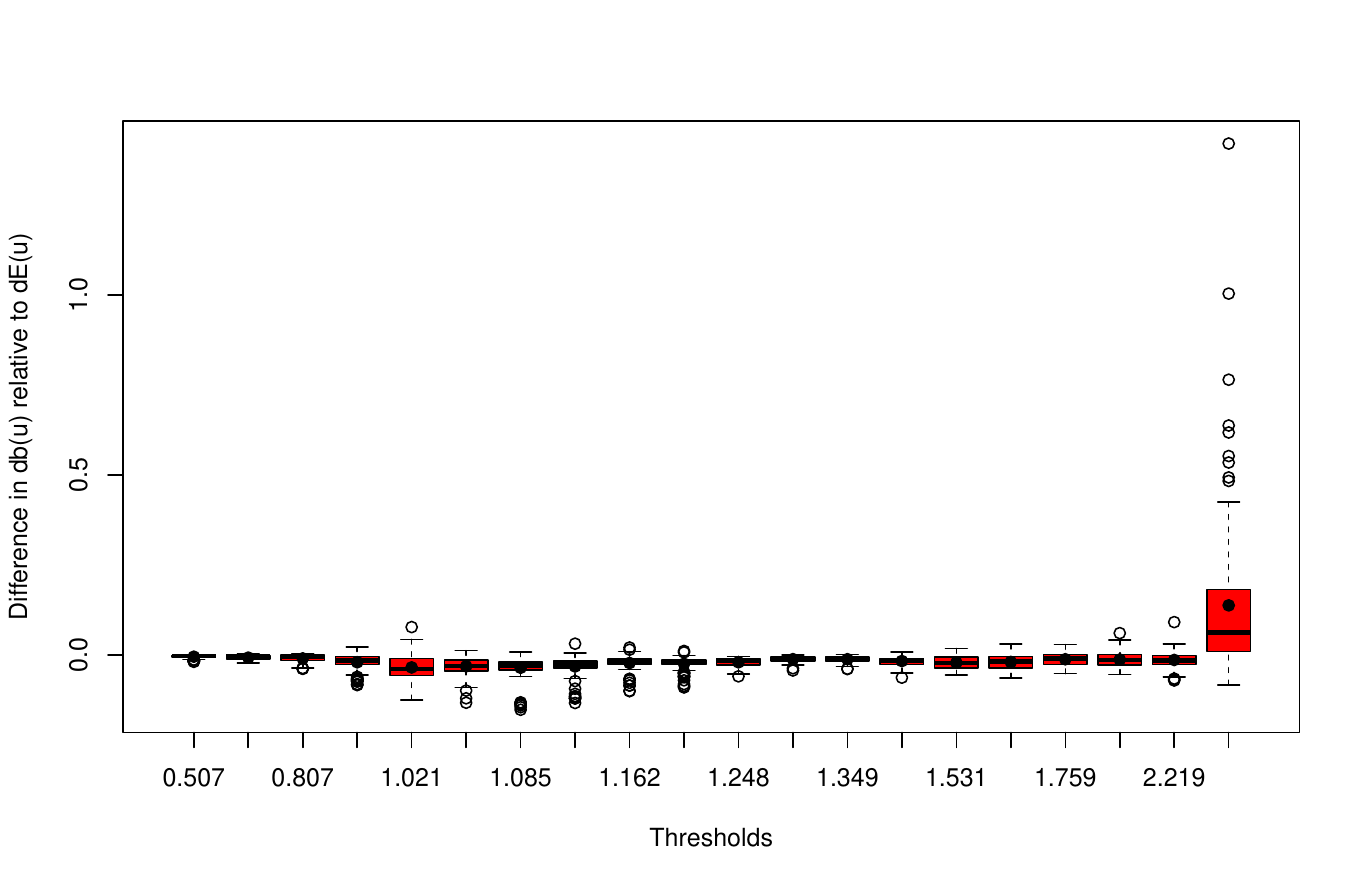}
    \caption{Plot of $(d^{500}_b(u)-d_b^{n_u}(u))/d_E(u)$ for first sample of Case 1. The mean for each threshold is shown as black points.}
    \label{fig: diff_dbu_rel_to_dEu}
\end{figure}

To demonstrate our findings from 
Figure~\ref{fig: m_500_vs_mnu} more concretely, Figure~\ref{fig: diff_dbu_rel_to_dEu} shows the sampling distribution of the difference $d^{500}_b(u) - d^{n_u}_b(u)$ relative to the metric value $d_E(u)$ for each threshold $u$. Across almost all thresholds, most of the values within the bootstrap sampling distribution lie very close to zero and more importantly, the mean of the sampling distribution lies very close to zero. These findings indicate that the choice of interpolation grid has no meaningful effect on the value of the metric for a particular threshold, in general. However, for the very largest threshold shown in the plot, we see a much larger range for the bootstrapped distribution of these relative differences, showing the effect of the over-sampling of the upper tail as you mentioned in your original review. Importantly, referring back to Figure~\ref{fig: m_500_vs_mnu}, the $d_b(u)$ values for the highest threshold are all larger than the largest $d_b(u)$ value for the optimal threshold in both plots. Thus, while there is a clear effect on the values of $d^{500}(u)$ and $d^{n_u}(u)$ for particular bootstrap samples, both Figure~\ref{fig: m_500_vs_mnu} and \ref{fig: diff_dbu_rel_to_dEu} show that the effect on the mean is relatively small and certainly, would not alter the selected threshold in any way. 

For Case 1, the true threshold is at a low sample quantile (16.67\%) and so, it is unlikely that the threshold choice would be affected by the under- or over-sampling of the tail as the only significant effect of this comes at very high thresholds. In a case where the optimal threshold lies at a higher sample quantile, based on the above analysis, we might expect the threshold choice to show greater sensitivity to choice of interpolation grid. To explore a case of this nature, we repeat the above analysis on the first sample from the replicated data of the Gaussian case. Here, we expect the optimal threshold to lie further into the tail due to the slow convergence of the Gaussian distribution to an extreme value limit. As a result, in theory, we expect that this case could show more sensitivity to the under- or over-sampling of the upper tail.

\begin{figure}[h]
    \centering
    \includegraphics[width=0.95\textwidth]{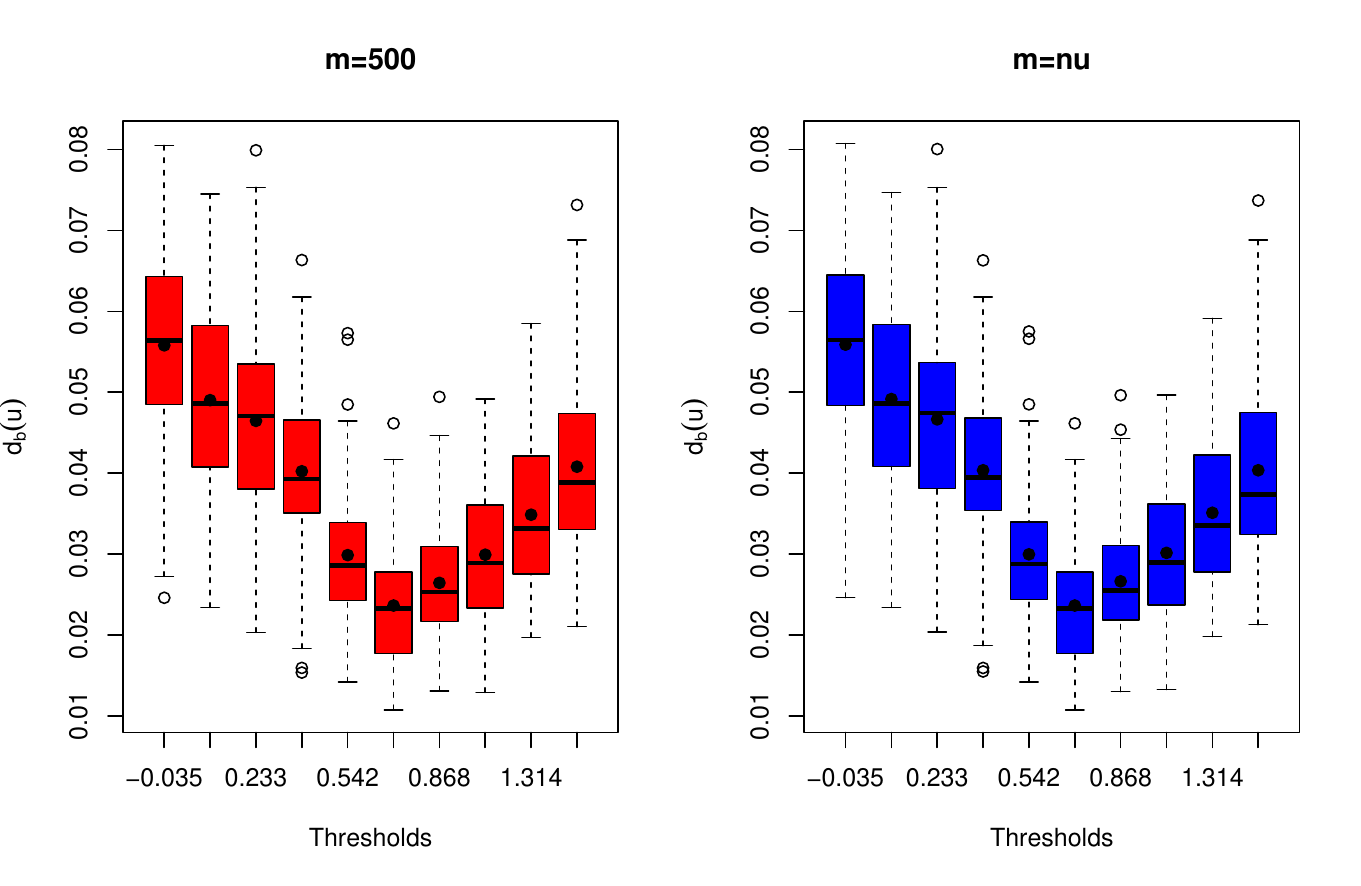}
    \caption{Boxplots of $d_b(u)$ for the first sample of the Gaussian case with $m=500$ (left) and $m=n_u$ (right). The mean for each threshold is shown as black points.}
    \label{fig: m_500_vs_mnu_gauss}
\end{figure}

\begin{figure}[H]
    \centering
    \includegraphics[width=0.95\textwidth]{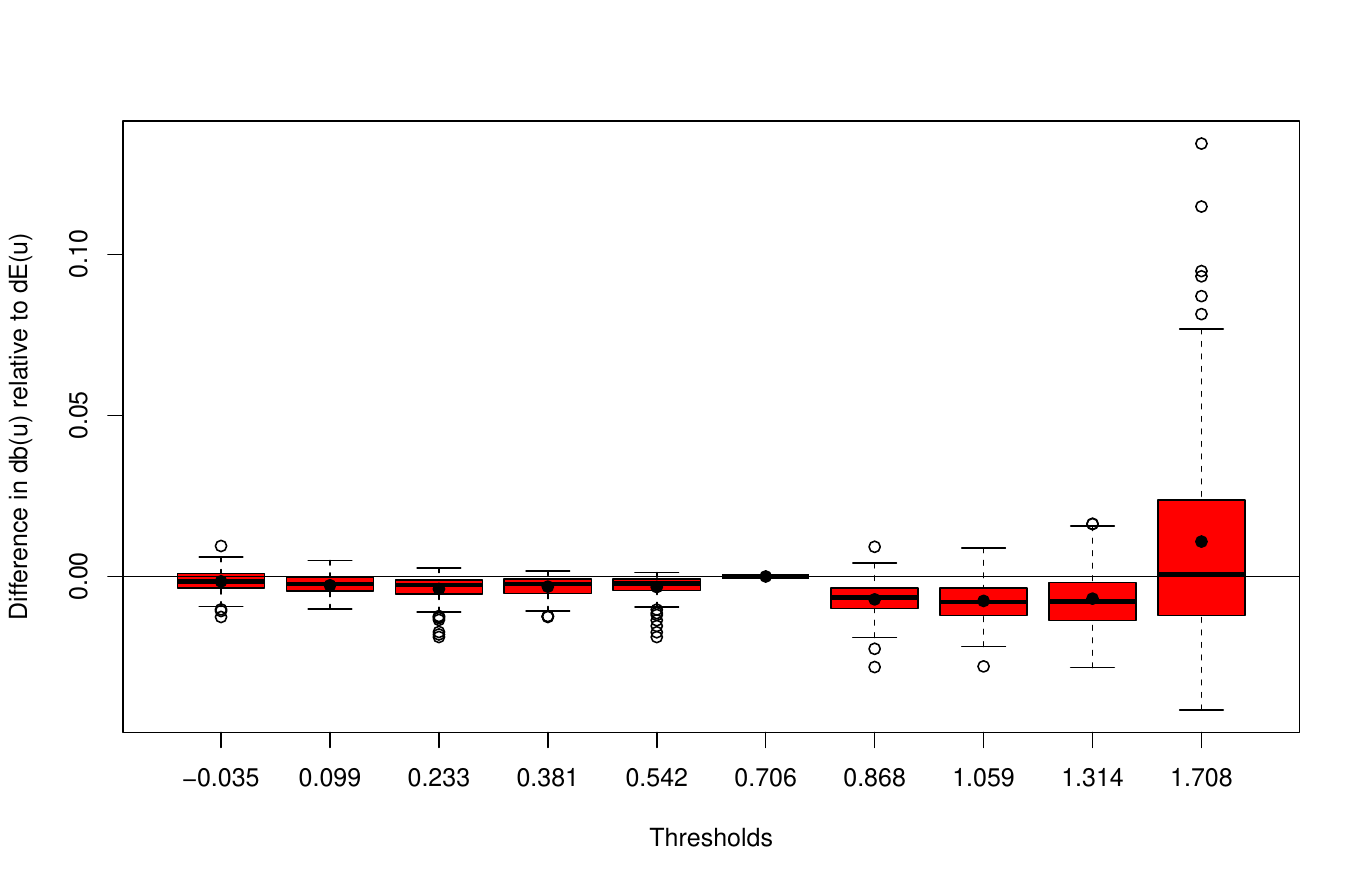}
    \caption{Plot of $(d^{500}_b(u)-d_b^{n_u}(u))/d_E(u)$ for first sample of the Gaussian case. The mean for each threshold is shown as black points.}
    \label{fig: diff_dbu_rel_to_dEu_gauss}
\end{figure}

Figures~\ref{fig: m_500_vs_mnu_gauss} and \ref{fig: diff_dbu_rel_to_dEu_gauss} show the Gaussian results, in the same format as for Case 1. In Figure~\ref{fig: m_500_vs_mnu_gauss}, there is a clear threshold choice in both plots and there does not seem to be any clear effect from the choice of interpolation grid on the bootstrap sampling distribution of mean-absolute deviations, and certainly there is no clear effect on the mean values for any thresholds. In fact, surprisingly, any effect due to under- or over-sampling the upper tail is much smaller in this case than above. Figure~\ref{fig: diff_dbu_rel_to_dEu_gauss} reiterates this where the sampling distribution of relative differences lie close to zero for all thresholds. The choice of interpolation grid does not show any meaningful effect on the overall metric value and certainly, would not affect the choice of threshold for this dataset. 

Given that our overall aim is to approximate the IAE between model quantiles and quantiles of the data generating process, we choose to keep the quality of the approximation of the IAE for particular samples fixed for a fair comparison across a range of candidate thresholds. Since $m$ controls the quality of the approximation, we suggest $m=500$ is sufficient based on the above results. Given we are confident in the quality of results for this choice, for convenience, we keep this fixed value across sample sizes and threshold levels in our simulation studies. 

In conclusion, in order to balance accuracy and computation time for our large scale simulation studies, we find that the EQD method, with value $B=100$ and $m=500$, provides sufficiently accurate results in a timely manner, so we choose to use these value throughout the paper, unless stated otherwise.

\subsection{Selection of calibration data in the metric $d_b(u)$}
\label{subsec:calibrationdata}

Here, we provide results for an adjusted version of the EQD method (as suggested by a referee) with the findings being reported in the main paper in Section~\ref{section:whyEQDchoice}. Here, the adjusted version takes
the calibration data used for threshold estimation as the actual observed sample excesses $\bm{x}_u$ of candidate threshold $u$, as described below. This adjusted version differs from the EQD method as proposed in Section~\ref{subsec:metric} of the main paper, where the calibration data for a particular bootstrapped sample are calculated based on the bootstrapped excesses $\bm{x}^b_u$. Thus, the only difference between the two approaches is that in $d_b(u)$, given by metric~\eqref{eqn: d_i}, the $Q(p_j; \bm{x}^b_u, \bm{q})$ term is replaced by  $Q(p_j; \bm{x}_u, \bm{q})$ in the adjusted version.

Table~\ref{tab: Qpx_vs_Qpxb_Case1-4} shows the RMSE, bias and variance of threshold estimates using the proposed and the adjusted versions of the EQD method evaluated on 500 replicated samples from each of Cases 1-4.
The relative performance of the two methods vary only slightly across all cases, and it is difficult to distinguish between the two methods, with each having the smaller RMSE an equal number of times. Furthermore,  any difference between the two methods is
very small relative to the differences between the EQD and the existing automated threshold selection approaches. 

Using the adjusted method leads to a larger variability in the distribution of $d_b(u)$ relative to the proposed method. This is because, unlike the proposed EQD method, the adjusted method does not compare like with like; one term in $d_b(u)$ is based on a bootstrap sample and the other on the actual data. Despite this increased variability, our simulation analysis suggests that the adjustment does not necessarily lead to a notable change in either the $d_E(u)$ value for a particular candidate threshold or in the subsequently selected threshold. As a result, we choose to retain our proposed method and this is utilised throughout the analyses of the main text and the supplementary material. 
\begin{table}[h!]
    \centering
    \begin{tabular}{|c|c|c|c||c|c|c|}
        \hline
        & \multicolumn{3}{|c||}{Proposed - $Q(p_j; \bm{x}^b_u, \bm{q})$} & \multicolumn{3}{|c|}{Adjusted - $Q(p_j; \bm{x}_u, \bm{q})$} \\
        \hline
        Case & RMSE & Bias & Variance & RMSE & Bias & Variance\\
        \hline
        Case 1 & 0.048 & $\bm{0.034}$ & 0.001 & $\bm{0.047}$ & 0.035 & 0.001 \\
        Case 2 & 0.060 & 0.031 & 0.003 & $\bm{0.050}$ & $\bm{0.027}$ & $\bm{0.002}$ \\
        Case 3 & $\bm{0.060}$ & $\bm{0.042}$ & 0.002 & 0.062 & 0.046 & 0.002 \\
        Case 4 & $\bm{0.526}$ & $\bm{-0.515}$ & 0.012 & 0.545 & $-0.535$ & $\bm{0.011}$ \\        \hline
    \end{tabular}
    \caption{RMSE, bias and variance of threshold estimates for Cases 1-4 for the EQD  method found using the different calibration data in the metric $d_b(u)$: the proposed EQD with $Q(p_j; \bm{x}^b_u, \bm{q})$ and the adjusted version with $Q(p_j; \bm{x}_u, \bm{q})$.}
    \label{tab: Qpx_vs_Qpxb_Case1-4}
\end{table}

\subsection{Investigating the effect of bootstrapping}\label{section: bootstrapping}
In this section, we provide further simulation experiments to investigate the effect of the bootstrapping component of the EQD method. We utilise a variant of the EQD method with no bootstrapping and compare the results against the original EQD method across Cases 0-4 and the Gaussian case (with $n=2000$), with each case based on 500 replicated samples (Case 0 is outlined in Section~\ref{section: furthercases}, while Cases 1-4 and the Gaussian case are outlined in Section~\ref{section: simstudy} of the main paper.).

Table~\ref{tab: noboot} provides results for the RMSE, bias and variance of threshold choices for the EQD method with and without bootstrapping, for Cases 0-4. For Cases 0-3, removing the bootstraps and only evaluating the metric on the original sample leads to RMSEs of threshold choice almost twice as large than for the original EQD method. In each of these cases, there is an increase in positive bias and the variance increases by a factor of at least 4. Thus, removing the bootstrapping component leads to higher and more variable threshold choices for these cases as would be expected. For Case 4, the removal of the bootstrapping component actually leads to a slight decrease in RMSE, in contrast to the previous cases. This decrease stems from a reduction in the negative bias component which is as a result of threshold choices being slightly higher than the original method. This reduction in bias comes at a cost of greater variability in selected thresholds. This greater variability is to be expected by not averaging over bootstrap samples.

\begin{table}[h]
    \centering
    \begin{tabular}{|c|c|c|c|c|c|c|c|c|c|}
    \hline
     & \multicolumn{3}{c|}{Case 0} & \multicolumn{3}{c|}{Case 1} & \multicolumn{3}{c|}{Case 2}\\
        \hline
         & RMSE & Bias & Variance & RMSE & Bias & Variance & RMSE & Bias & Variance \\
        \hline
        Original & 0.042 & 0.019 & 0.001 & 0.048 & 0.034 & 0.001 & 0.060 & 0.031 & 0.003\\
        No bootstraps & 0.095 & 0.039 & 0.008 & 0.090 & 0.054 & 0.005 & 0.122 & 0.063 & 0.011\\
        \hline
    \end{tabular}
    
        \vspace{0.5cm} 
    \begin{tabular}{|c|c|c|c|c|c|c|}
    \hline
     & \multicolumn{3}{c|}{Case 3} & \multicolumn{3}{c|}{Case 4}\\
        \hline
        & RMSE & Bias & Variance & RMSE & Bias & Variance\\
        \hline
        Original & 0.060 & 0.042 & 0.002 & 0.526 & $-0.515$ & 0.012\\
        No bootstraps & 0.138 & 0.080 & 0.013 & 0.473 & $-0.441$ & 0.030\\
        \hline
    \end{tabular}
    \caption{RMSE, bias and variance of threshold choice for Cases 0-4 for EQD only evaluated on original sample (i.e., no bootstrapping).}
    \label{tab: noboot}
\end{table}

Table~\ref{tab: gauss_noboot} shows the RMSE of quantile estimation following threshold selection using the original EQD method and the variant with no bootstrapping for the Gaussian case. There is a slight improvement in performance in terms of RMSE from including the bootstrapping component across all quantiles but results are very similar across both methods. 

\begin{table}[h]
    \centering
    \begin{tabular}{|c|c|c|}
    \hline
     $j$ & Original & No bootstraps \\
     \hline
     0 & 0.214 & 0.217 \\
     1 & 0.430 & 0.435  \\
     2 & 0.703 & 0.715  \\
     \hline
    \end{tabular}
    \caption{RMSEs of estimated $(1-p_{j,n})$-quantiles where $p_{j,n}=1/(10^jn)$, for $j=0,1,2$, for the Gaussian case.}
    \label{tab: gauss_noboot}
\end{table}

Overall, across the studied cases, the addition of bootstrapping to the EQD method leads to a systematic reduction in the variance and lower threshold choices, typically resulting in reduced RMSE, for the selected thresholds and the subsequent quantile estimates.  While the mean absolute deviation is a robust and effective metric for threshold selection, the bootstrapping component provides further stability in the threshold choices and allows us to account for the increasing uncertainty in parameter estimates as the threshold increases. These investigations support our choice to use bootstrapping as a key component of the EQD method.

\section{More simulation study results linked to  Section~\ref{section: simstudy}}\label{section: extra_sim_res}

\subsection{Overview}

We provide results that expand on those in Section~\ref{section: simstudy} of the main text. 
Section~\ref{section: detailed_sim_res} provides bias-variance decompositions for the RMSEs values given in Section~\ref{section: simstudy} and presents additional results for the \citet{danielsson2001using,danielsson2019} methods, evaluating threshold estimation for Cases 1-4 and quantile estimation for Gaussian data. In Section~\ref{section: furthercases}, we provide further threshold estimation results for additional cases, which have alternative parameters and sample sizes to those of Cases 1-4. Section~\ref{section: mode_candidate_grids} presents additional results to assess the sensitivity of the methods to the choice of candidate threshold grids using Cases 1 and 4 with candidate grids defined above the mode of the distribution and for the Gaussian case with candidate grids spanning the entire range of the sample.

\subsection{Detailed results for the case studies}
\label{section: detailed_sim_res}

We provide the RMSE and bias-variance decomposition for the application of the EQD method and all methods described in Section~\ref{section: current} of the main text to the case studies detailed in Section~\ref{section: simstudy} of the main text.

In what follows, we find here that the \citet{danielsson2001using,danielsson2019} methods
perform much worse that the EQD, \citet{wadsworth2016exploiting} and \citet{northrop2017cross} methods both in terms of threshold and quantile estimation. We therefore omit results for these methods in Section~\ref{section: simstudy} of the main text and do not to apply them beyond this section of the supplementary material.

\subsubsection{Scenario 1: True GPD tail - Cases 1-4}

\textbf{Threshold recovery:}\ \\
For Cases 1-4, Table~\ref{tab: biasvarcases} shows the RMSE, bias and variance of the thresholds selected by the EQD, \citet{wadsworth2016exploiting} and \citet{northrop2017cross} methods.
The RMSE is also reported in Table~\ref{tab: rmse_thr} of the main paper, 
but is repeated here for completeness. The 
EQD method has the smallest RMSE and the least variable estimates in all cases, and is the least biased method Cases 1-3.  
Table~\ref{tab: biasvarcases} also presents equivalent results for the \citet{danielsson2001using, danielsson2019} methods. Both methods show considerably larger RMSEs than the EQD, \citet{wadsworth2016exploiting} and \citet{northrop2017cross} methods, due to the large positive biases of these methods across all cases. In particular, for Cases 3 and 4, the \citet{danielsson2019} method has the smallest variance of all the methods but its larger bias leads to RMSE values much larger than those of the EQD and other methods in Table~\ref{tab: biasvarcases}. 

\begin{table}[h!]
    \centering
    \begin{tabular}{|c|c|c|c||c|c|c||c|c|c|}
        \hline
        & \multicolumn{3}{|c||}{\textit{EQD}} & \multicolumn{3}{|c||}{\textit{Wadsworth}\tablefootnote{Results for Wadsworth are calculated only on the samples where a threshold was estimated. It failed estimate a threshold for 2.4\%, 26.4\%, 0\%, 3.6\% of the simulated samples in Cases 1-4, respectively. \label{note_wads_fail_sm_1}}} & \multicolumn{3}{|c|}{\textit{Northrop}} \\
        \hline
        Case & RMSE & Bias & Variance & RMSE & Bias & Variance & RMSE & Bias & Variance\\
        \hline
        Case 1 & $\bm{0.048}$ & $\bm{0.034}$ & $\bm{0.001}$ & 0.349 & 0.111 & 0.110 & 0.536 & 0.276 & 0.212 \\
        Case 2 & $\bm{0.060}$ & $\bm{0.031}$ & $\bm{0.003}$ & 0.461 & 0.204 & 0.172 & 0.507 & 0.238 & 0.201 \\
        Case 3 & $\bm{0.060}$ & $\bm{0.042}$ & $\bm{0.002}$ & 0.221 & 0.060 & 0.045 & 0.463 & 0.256 & 0.149 \\
        Case 4 & $\bm{0.526}$ & $-0.515$ & $\bm{0.012}$ & 0.627 & $-0.407$ & 0.230 & 0.543 & $\bm{-0.222}$ & 0.246 \\
        \hline
    \end{tabular}
\vspace{0.5cm}

    \begin{tabular}{|c|c|c|c||c|c|c|}
        \hline
        & \multicolumn{3}{|c||}{\textit{\citet{danielsson2001using}}} &  \multicolumn{3}{|c|}{\textit{\citet{danielsson2019}}} \\
        \hline
        Case & RMSE & Bias & Variance & RMSE & Bias & Variance \\
        \hline
        Case 1 & 2.767 & 2.416 & 1.825  & 1.635 & 1.633 & 0.007 \\
        Case 2 & 2.212 & 1.850 & 1.474  & 1.639 & 1.634 & 0.017 \\
        Case 3 & 2.528 & 2.441 & 0.435 & 1.314 & 1.314 & 0.001 \\
        Case 4 & 2.838 & 2.499 & 1.813 & 1.138 & 1.134 & 0.009 \\
        \hline
    \end{tabular}
    \caption{RMSE, bias and variance of the threshold estimates for Cases 1-4: for EQD, Wadsworth and Northrop methods (top) and 
    \citet{danielsson2001using, danielsson2019} (bottom).   Results are based on 500 replicated samples.}
    \label{tab: biasvarcases}
\end{table}

\FloatBarrier

\textbf{Quantile recovery:}\\
Tables~\ref{tab: bias_quant} \& \ref{tab: var_quant} present the bias and variance of the quantile estimates for the EQD, \citet{wadsworth2016exploiting} and \citet{northrop2017cross} methods applied to Cases 1-4. These are shown for the $(1-p_{j,n})$-quantile estimates where $p_{j,n} = 1/(10^jn)$ for $j=0,1,2$ and $n$ denotes the length of the simulated dataset. As mentioned in Section~\ref{section: simstudy} of the main paper, we use exceedance probabilities of this form because we have simulated samples of different sizes and want to make extrapolation equally difficult in each case. These bias and variance components correspond to the RMSE values presented in Table~\ref{tab: rmse_quant} in Section~\ref{section: simstudy} of the main paper, where the EQD method achieves the lowest RMSEs in all cases and quantiles. 
Table~\ref{tab: var_quant} shows that these lower RMSE values derive mainly from the variance component; for all $j$ and in all cases the EQD method shows the least variability in quantile estimates, with the differences between the methods becoming more evident for higher $j$. The smallest absolute bias values in Table~\ref{tab: bias_quant} vary between each of the methods but the EQD incurs the least bias in the majority of cases and quantiles. In particular, the EQD achieves the smallest absolute bias values for all $j$ in Cases 2 and 3.
\begin{table}[h!]
    \centering
    \begin{tabular}{ |c|c|c|c||c|c|c| } 
    \hline 
     & \textit{EQD} & \textit{Wadsworth\footref{note_wads_fail_sm_1}} & \textit{Northrop} &  \textit{EQD} & \textit{Wadsworth\footref{note_wads_fail_sm_1}} & \textit{Northrop} \\
     \hline 
     $j$ & \multicolumn{3}{|c||}{\textbf{Case 1}} & \multicolumn{3}{|c|}{\textbf{Case 2}} \\
     \hline
     0 & $\bm{-0.021}$ & $-0.079$  &  $-0.075$ &  $\bm{-0.049}$ & $-0.118$ & $-0.071$ \\ 
     1 & $-0.015$ & $-0.192$ & $\bm{-0.001}$ &  $\bm{-0.046}$ & $-0.316$ & 0.245 \\ 
     2 & $\bm{0.044}$ & $-0.319$ & 0.554 &  $\bm{0.069}$ & $-0.532$ & 2.568 \\       
     \hline
     \hline
    & \multicolumn{3}{|c||}{\textbf{Case 3}} & \multicolumn{3}{|c|}{\textbf{Case 4}} \\
     \hline
     0 & $\bm{-0.008}$ & $-0.026$  &  $-0.041$ & $-0.283$ & $-0.372$ & $\bm{-0.192}$ \\ 
     1 & $\bm{-0.007}$ & $-0.047$ & $-0.065$ &  $-0.722$ & $-0.965$ & $\bm{-0.344}$ \\ 
     2 & $\bm{-0.002}$ & $-0.066$ & $-0.074$ &  $-1.410$ & $-1.809$ & $\bm{-0.258}$ \\       
     \hline
    \end{tabular}
\caption{Bias of the estimated quantiles in Cases 1-4 based on fitted GPD above chosen threshold. The smallest absolute bias for each quantile are highlighted in bold.}
 \label{tab: bias_quant}
\end{table}

\begin{table}[h!]
    \centering
    \begin{tabular}{ |c|c|c|c||c|c|c| } 
    \hline 
     & \textit{EQD} & \textit{Wadsworth\footref{note_wads_fail_sm_1}} & \textit{Northrop} &  \textit{EQD} & \textit{Wadsworth\footref{note_wads_fail_sm_1}} & \textit{Northrop} \\
     \hline 
     $j$ & \multicolumn{3}{|c||}{\textbf{Case 1}} & \multicolumn{3}{|c|}{\textbf{Case 2}} \\
     \hline
     0 & $\bm{0.317}$ & 0.348 & 0.565 &  $\bm{0.358}$ & 0.386 & 0.538 \\ 
     1 & $\bm{1.585}$ & 1.903 & 5.657 &  $\bm{2.215}$ & 2.611 & 12.305 \\ 
     2 & $\bm{6.000}$ & 7.297 & 50.155 &  $\bm{9.743}$ & 11.885 & 519.569 \\       
     \hline
     \hline
    & \multicolumn{3}{|c||}{\textbf{Case 3}} & \multicolumn{3}{|c|}{\textbf{Case 4}} \\
     \hline
     0 & $\bm{0.036}$ & 0.038  &  0.051 & $\bm{0.379}$ & 0.503 & 0.591 \\ 
     1 & $\bm{0.105}$ & 0.116 & 0.199 &  $\bm{1.926}$ & 3.317 & 4.805 \\ 
     2 & $\bm{0.233}$ & 0.263 & 0.549 &  $\bm{7.287}$ & 16.879 & 30.999 \\       
     \hline
    \end{tabular}
\caption{Variance of the estimated quantiles in Cases 1-4 based on fitted GPD above chosen threshold. The smallest variance for each quantile are highlighted in bold.}
 \label{tab: var_quant}
\end{table}
\FloatBarrier

For completeness, Table~\ref{tab: RMSE_quant_dan} presents the equivalent RMSE values for the \citet{danielsson2001using,danielsson2019} methods. These can be compared with the RMSE results for the EQD, \citet{wadsworth2016exploiting} and \citet{northrop2017cross} methods in Table~\ref{tab: rmse_quant} of the main paper.
For $j=0$ the threshold choice should not be too important because we are not extrapolating but Table~\ref{tab: RMSE_quant_dan} shows that, even in this case, the \citet{danielsson2001using, danielsson2019} approaches have RMSE values much greater than the other methods, by factors of between $1.5-3$ across the cases. This difference in performance is only exacerbated as we extrapolate further. For $j=1,2$, both approaches lead to RMSEs that are orders of magnitude larger than any of the other methods analysed. For example, when $j=2$ the RMSEs of the \citet{danielsson2001using} and \citet{danielsson2019} methods are respectively $4-70$ and $3-7$ times larger than those of the EQD method.

\begin{table}[h!]
    \centering
    \begin{tabular}{|c|c|c||c|c|} 
    \hline 
    $j$ & \textit{Danielsson 2001} & \textit{Danielsson 2019} &  \textit{Danielsson 2001} & \textit{Danielsson 2019} \\
     \hline 
     & \multicolumn{2}{|c||}{\textbf{Case 1}} & \multicolumn{2}{|c|}{\textbf{Case 2}} \\
     \hline
     $0$ & 1.020 & 0.859  &  0.962 &  0.757 \\ 
     $1$ & 3.128 & 3.172 & 3.806 &  3.991 \\ 
     $2$ & 12.943 & 10.303 & 110.347 &  23.102 \\       
     \hline
     \hline
    & \multicolumn{2}{|c||}{\textbf{Case 3}} & \multicolumn{2}{|c|}{\textbf{Case 4}} \\
     \hline
     $0$ & 0.675 & 0.262 & 1.118 & 0.865 \\
     $1$ & 1.655 & 0.570 &  3.938 & 2.978 \\ 
     $2$ & 38.001 & 1.000 & 40.653 & 8.721  \\       
     \hline
    \end{tabular}
\caption{RMSEs in the estimated quantiles in Cases 1-4 based on fitted GPD above chosen threshold for the \citet{danielsson2001using} and \citet{danielsson2019} methods.}
 \label{tab: RMSE_quant_dan}
\end{table}

\subsubsection{Scenario 2: Gaussian data}

\textbf{Quantile recovery:}\\
We next consider the case of Gaussian data, using 500 simulated datasets of size $n=2000$ and $20000$. Tables~\ref{tab: gaussianRMSE_bias_var} and \ref{tab: gaussianRMSE_bias_var_large} present the bias, variance and RMSE for the estimation of the $(1-p_{j,n})$-quantiles (where $p_{j,n}=1/(10^jn)$ and $j=0,1,2$), based on the thresholds selected by the EQD, \citet{wadsworth2016exploiting} and \citet{northrop2017cross} methods. These RMSE values are detailed in Table~\ref{tab: rmse_norm} of the main paper.

Table~\ref{tab: gaussianRMSE_bias_var} shows that for the smaller sample size of $n=2000$, the EQD method achieves the smallest RMSEs and variance for all $j$ for all methods. The \citet{danielsson2019} and \citet{northrop2017cross} methods incur the least absolute bias in quantile estimation due to their slightly higher threshold choices, see Table~\ref{tab: norm_thresh_choices}, and have similar RMSE values smaller than that of the \citet{wadsworth2016exploiting} method in this aspect. The \citet{danielsson2001using} method incurs considerable bias with large variance in its quantile estimates, leading to the highest RMSEs of all analysed methods.

\begin{table}[h!]
    \centering
    \begin{tabular}{|c|c|c|c|c|c|c|c|c|c|}
    \hline
    & \multicolumn{3}{|c|}{ \textit{EQD}} & \multicolumn{3}{|c|}{ \textit{Wadsworth}\tablefootnote{Results for the Wadsworth method, which failed on 0.4\% of the samples here, are calculated only for samples where a threshold estimate was obtained. \label{note_wads_fail_sm_2}}} & \multicolumn{3}{|c|}{\textit{Northrop}} \\
    \hline
     $j$ & RMSE & Bias & Variance & RMSE & Bias & Variance & RMSE & Bias & Variance \\
     \hline
     0 & $\bm{0.214}$ & $-0.086$ & $\bm{0.038}$ & 0.239 & $-0.120$ & 0.043 & 0.225 & $-0.076$ & 0.045  \\
     1 & $\bm{0.430}$ & $-0.275$ & $\bm{0.109}$ & 0.529 & $-0.366$ & 0.147 & 0.461 & $-0.224$ & 0.162  \\
     2 & $\bm{0.703}$ & $-0.521$ & $\bm{0.222}$ & 0.890 & $-0.654$ & 0.365 & 0.765 & $-0.414$ & 0.414  \\ 
     \hline
    \end{tabular}
    
    \vspace{0.5cm} 
 \begin{tabular}{|c|c|c|c|c|c|c|}
    \hline
    & \multicolumn{3}{|c|}{\textit{\citet{danielsson2001using}}} & \multicolumn{3}{|c|}{\textit{\citet{danielsson2019}}}  \\
    \hline
     $j$ & RMSE & Bias & Variance & RMSE & Bias & Variance \\
     \hline
     0 & 0.758 & $-0.470$ & 0.354 & 0.232 & $\bm{-0.059}$ & 0.050\\
     1 & 1.550 & $-0.815$ & 1.739 & 0.479 & $\bm{-0.173}$ & 0.200  \\
     2 & 33.183 & $-1.380$ & 1099.182 & 0.790 & $\bm{-0.321}$ & 0.522 \\
     \hline
    \end{tabular}
    \caption{RMSE, bias and variance of estimated quantiles from a Gaussian distribution with sample size of $2000$: for EQD, Wadsworth and Northrop methods (top) and 
    \citet{danielsson2001using, danielsson2019} (bottom).   Results are based on 500 replicated samples.}
    \label{tab: gaussianRMSE_bias_var}
\end{table}

As the sample size $n$ increases, the relative importance of bias and variance terms within the RMSE shifts, with low bias becoming increasingly important. Table~\ref{tab: gaussianRMSE_bias_var_large} 
shows that when $n=20000$, the \citet{northrop2017cross} method achieves the lowest RMSE and bias over all $j$. The EQD method takes second place, again showing the least variability in its estimates. Note that all of the methods show decreased RMSEs as $n$ increases from $2000$ to $20000$, even though the bias values do not all decrease. A possible reason for this lack of reduction in bias is the slow convergence of the Gaussian distribution to the extreme value limit, so an order of magnitude increase in sample sizes could be required for the bias to reduce. We attempted to apply the \citet{danielsson2001using, danielsson2019} methods to the 500 Gaussian samples with the larger sample size of $n=20000$ but the computation time was simply too large.

\begin{table}[h!]
    \centering
    \begin{tabular}{|c|c|c|c|c|c|c|c|c|c|}
    \hline
    & \multicolumn{3}{|c|}{ \textit{EQD}} & \multicolumn{3}{|c|}{ \textit{Wadsworth}} & \multicolumn{3}{|c|}{\textit{Northrop}} \\
    \hline
     $j$ & RMSE & Bias & Variance & RMSE & Bias & Variance & RMSE & Bias & Variance \\
     \hline
     0 & 0.187 & $-0.131$ & $\bm{0.018}$ & 0.214 & $-0.165$ & 0.019 & $\bm{0.172}$ & $\bm{-0.104}$ & 0.019  \\
     1 & 0.368 & $-0.307$ & $\bm{0.042}$ & 0.422 & $-0.366$ & 0.044 & $\bm{0.331}$ & $\bm{-0.255}$ & 0.045 \\
     2 & 0.594 & $-0.528$ & $\bm{0.074}$ & 0.672 & $-0.611$ & 0.078 & $\bm{0.533}$ & $\bm{-0.450}$ & 0.081 \\ 
     \hline
    \end{tabular}
    \caption{RMSE, bias and variance of estimated quantiles from a Gaussian distribution with sample size of $20000$. Results are based on 500 replicated samples.}
    \label{tab: gaussianRMSE_bias_var_large}
\end{table}
\FloatBarrier

\textbf{Threshold Recovery:} \\
There is no true GPD threshold for the Gaussian scenario but 
the quantile estimates discussed in 
Tables~\ref{tab: gaussianRMSE_bias_var} and \ref{tab: gaussianRMSE_bias_var_large} require a preceding step of selecting a suitable threshold above which the GPD approximation is adequate. In Table~\ref{tab: norm_thresh_choices}, we provide information about the selected thresholds for the Gaussian data, presenting the 2.5\%, 50\%, 97.5\% values of the sampling distribution of the threshold estimates (presented as a quantile of the Gaussian distribution for each method). 
The results show that the \citet{wadsworth2016exploiting} method tends to estimate the threshold lowest, followed by the EQD method, and then the \citet{northrop2017cross} method which tends to estimate the highest threshold values, with this finding being consistent across the sampling distribution quantiles. It is interesting to see that, even with a very large sample size, almost always thresholds are estimated to be below the 95\% quantile (the maximum of the candidate thresholds) of the Gaussian distribution. This is surprising given its widely known slow convergence issues. 

\begin{table}[ht]
    \centering
    \begin{tabular}{|c|c|c|c||c|c|c|}
    \hline
    & \multicolumn{3}{|c||}{$n=2000$} & \multicolumn{3}{|c|}{$n=20000$} \\
    \hline
      & \textit{EQD} & \textit{Wadsworth}\footref{note_wads_fail_sm_2} & \textit{Northrop} & \textit{EQD} & \textit{Wadsworth} & \textit{Northrop} \\
     \hline
     50\% $Q$ & 75 & 50 & 80 & 87.5 & 84 & 91.5 \\
     2.5\%, 97.5\% $Q$ & 55, 90 &  50, 95 &  60, 95 &  77.5, 94 &  69.5, 95 &  82.5, 95  \\
     \hline
    \end{tabular}
    \caption{Sampling distribution quantiles (2.5\%, 50\%, 97.5\%) 
    of quantile level $Q$ (\%) for the threshold estimates for each method derived from 500 replicated samples from a Gaussian distribution for two sample sizes $n$.}
    \label{tab: norm_thresh_choices}
\end{table}

\subsection{Extra case studies}\label{section: furthercases}

This section provides a description and the results of additional case studies, beyond Cases 1-4, which were omitted from Section~\ref{section: simstudy} in the main paper. 
These are denoted by Cases 0, 5, 6, 7 and 8. Specifically, these extra cases are:\\
\textit{Case 0}: We simulate samples of size $n=1000$ from a GPD(0.5,0.1) above a threshold $u=1$ with no data below the threshold.\\
\textit{Case 5}: We simulate samples from the 
distribution~\eqref{eqn: case1-3cdf} with $(\sigma_u, \xi) = (0.5, 0.1)$, but with a reduced sample size of $n=120$.\\ 
\textit{Case 6 \& 7}: We simulate samples from the 
distribution~\eqref{eqn: case1-3cdf} with a sample size of $n=1200$, but with shape parameters $\xi < -0.05$, i.e., $\xi = -0.2$ for Case 6 and $\xi = -0.3$ for Case 7.\\
\textit{Case 8}: We simulate samples from the 
distribution~\eqref{eqn: case1-3cdf} with $(\sigma_u, \xi) = (0.5, 0.1)$, but with an increased sample size of $n=20000$.

We chose to omit Case 0 from the main text due to its simplicity, in that it should be the easiest case for threshold selection due to the lack of data below the threshold. Cases~5, 6 and 7 were also omitted due to poor performance of the \citet{wadsworth2016exploiting} method which failed to estimate a threshold in the majority of samples generated from each of these cases. 
This high failure rate occurs for two reasons; dependence between parameter estimates when using candidate thresholds which lie in close proximity for small sample sizes or in the cases where $\xi < -0.05$, where an error results from a divergent integral in the calculation of the inverse Fisher information matrix. Thus, for Cases 5, 6 and 7, here we only show comparisons of the EQD method against the \citet{northrop2017cross} method. Finally, Case 8 was also omitted from the main text due to the large sample size being atypical of extreme value analyses, but we decided to include a large sample case to explore how the EQD compares with existing methods which base their theoretical justification on asymptotic arguments.

In all of the cases, results presented here, for each of the methods, are based on 500 replicated samples and we take the set of candidate thresholds as the sample quantiles at levels 0\%, 5\% \ldots, 95\%, as in Section~\ref{section: simstudy} of the main paper. Specifically, for Case 8, we considered an additional finer grid of candidate thresholds as the sample quantiles at levels 0\%, 0.5\%, \ldots, 95\%. For Case 0, we initially used a set of candidate thresholds which contains values lower than the minimum of the sample, but \citet{wadsworth2016exploiting} and \citet{northrop2017cross} methods had major problems, and so we omit those results. However, this restriction to consider only candidate thresholds which are sample quantiles automatically positively biases threshold estimates in Case 0.
In any simulation study, it is reasonable to consider candidate thresholds above and below the true threshold, so the fact that the EQD method continues to work well for candidate thresholds below the sample quantiles is a particularly pleasing feature, even if not illustrated here. In Section~\ref{section: mode_candidate_grids}, we explore the effect of candidate thresholds which lie below the mode of the data for Cases 1, 4 and the Gaussian case for the EQD, Wadsworth and Northrop methods.

Case 0 should be much easier to estimate than for even Cases 1-3,
with the lowest candidate threshold being very close to the true threshold. Table~\ref{tab: RMSE_Case0} provides the RMSE, bias and variance of threshold estimates for each of the methods for this case, with the EQD obtaining the lowest value for each of the three summary features  by considerable margins. Hence, in the most ideal case for threshold estimation, the EQD method excels in its performance.

\begin{table}[h!]
    \centering
    \begin{tabular}{|c|c|c|c|} 
    \hline
     & \textit{EQD} & \textit{Wadsworth\tablefootnote{Results for Wadsworth are calculated only on the samples where a threshold was estimated, the method failed on 3.8\% of the simulated samples in Case 0. \label{note_wads_fail_sm_0i}}} & \textit{Northrop} \\
     \hline
      RMSE & $\bm{0.042}$ & 0.564  &  0.566 \\ 
      Bias & $\bm{0.019}$ & 0.228 & 0.326  \\ 
      Variance & $\bm{0.001}$ & 0.266 & 0.214  \\ 
     \hline
    \end{tabular}
\caption{RMSE, bias and variance of the threshold estimates for Case 0. Results are based on 500 replicated samples. The smallest in each case is given in bold.}
 \label{tab: RMSE_Case0}
\end{table}

Table~\ref{tab: RMSE_567} shows the RMSEs for the EQD and \citet{northrop2017cross} methods for estimating the threshold in Case 5, 6 and 7, with the EQD performing the best on each occasion. Case 5 is particularly important as the small sample size 
is typical of many data applications, so it is especially pleasing to see that the EQD method outperforms the \citet{northrop2017cross} method by the largest of the three margins in this case. \citet{northrop2017cross} performs especially badly in Case 5 in terms of variance. Furthermore, the EQD method, relative to that of \citet{northrop2017cross}, has a smaller (equal) absolute error in threshold estimates in 70.0\% (20.4\%) of samples. Similarly, in Cases 6 and 7, the EQD achieves a smaller (equal) absolute error in 59.8\% (18.0\%) of samples and 50.2\% (18.8\%) respectively.   

\begin{table}[h!]
    \centering
    \begin{tabular}{|c|c|c|c|} 
    \hline
     & Case 5 & Case 6 & Case 7 \\
     \hline
      \textit{EQD} & $\bm{0.078}$ & $\bm{0.107}$  &  $\bm{0.185}$ \\ 
      \textit{Northrop} & 0.602 & 0.373 & 0.341  \\ 
     \hline
    \end{tabular}
\caption{RMSEs of the threshold estimates of the EQD and Northrop methods in Cases 5, 6 and 7.
The smallest value in each case is given in bold.}
 \label{tab: RMSE_567}
\end{table}

Table~\ref{tab: rmse_thr_case_1_large} shows the RMSEs of threshold choice for Case 8. The results are shown for each method using two different candidate grids of thresholds. In contrast to the previous results, the Wadsworth method slightly outperforms the EQD achieving the smallest RMSEs for these large samples. However, the sample size for this to be achieved significantly exceeds that for data in practice. 
This illustrates the potential benefits, but also serious limitations, of relying on asymptotic methods to guide threshold selection.

\begin{table}[h!]
    \centering
    \begin{tabular}{|c||c||c||c|}
        \hline
        Grid (\% quantile) & \textit{EQD} & \textit{Wadsworth} & \textit{Northrop}\\
        \hline
         0 (5) 95 & 0.036  & $\bm{0.021}$  & 0.503  \\
        0 (0.5) 95 & 0.027  & $\bm{0.003}$  & 0.529  \\
        \hline
    \end{tabular}
    \caption{RMSE of the threshold choices for Case 8 for two different candidate grids given with notation \textit{start (increment) end}. The smallest values are in bold.}
    \label{tab: rmse_thr_case_1_large}
\end{table}

Table~\ref{tab: rmse_quant_Case_1_large} shows the RMSEs of quantile estimation for Case 8, where for threshold selection the Wadsworth method has a slight benefit over EQD. Here, for quantile estimation, there are similar findings, the Wadsworth method achieves the lowest RMSEs, followed closely by the EQD and then, the Northrop method which obtains significantly higher RMSEs. 

\begin{table}[h!]
    \centering
    \begin{tabular}{ |c|c|c|c||c|c|c| } 
    \hline 
    Grid & \multicolumn{3}{|c||}{0 (5) 95} & \multicolumn{3}{|c|}{0 (0.5) 95} \\
     \hline 
     $j$  & \textit{EQD} & \textit{Wadsworth} & \textit{Northrop} &  \textit{EQD} & \textit{Wadsworth} & \textit{Northrop} \\
     \hline
     0 & 0.370 & $\bm{0.364}$ & 0.546 &  0.369 & $\bm{0.358}$ & 0.573 \\ 
     1 & 0.694 & $\bm{0.681}$ & 1.157 &  0.692 & $\bm{0.669}$ & 1.200 \\ 
     2 & 1.199 & $\bm{1.174}$ & 2.189 &  1.196 & $\bm{1.151}$ & 2.242 \\       
     \hline
    \end{tabular}
\caption{RMSEs in the estimated quantiles in Case 8 samples based on fitted GPD above chosen threshold for two candidate grids given with notation \textit{start (increment) end}. The smallest RMSE for each quantile are highlighted in bold.}
 \label{tab: rmse_quant_Case_1_large}
\end{table}

\FloatBarrier

\subsection{Sensitivity to the choice of candidate threshold grids}\label{section: mode_candidate_grids} 
In this section, we provide further simulation experiments where we choose the candidate threshold grids more in line with general extreme value analyses (i.e., above the mode) and explore the sensitivity of methods to candidate thresholds which lie below the mode of the data. Since the GPD density is monotonically decreasing for realistic values of $\xi$ (i.e., when $\xi>-1$), when utilising a threshold selection procedure, it is unusual to consider thresholds which lie below an obvious mode. Many threshold selection procedures are simply not set up to handle data from non-monotonically decreasing densities. So we explore a different choice for the candidate threshold grids here to ensure that our choices made in the main paper do not unfairly favour the EQD method.

Given that our aim is to provide a method which requires no user input, we consider candidate thresholds across the range of the sample data and allow the method to make the threshold selection whether the conditions are suitable or not. In the main paper and the supplementary material, outside of Gaussian case where we specifically restricted our choice of candidate thresholds such that $u > q_{50}$ (with $q_{50}$ being the sample median), and Case 0, where the optimal threshold is at the lowest sample quantile, we have only explored cases where candidate thresholds span the range of the dataset. Here, we reanalyse two cases; Case 1 and Case 4, now with candidate threshold grids which lie above the mode of the distributions.

To avoid sensitivities from different mode estimators clouding the results, we use the true mode to help define our range/set of candidate thresholds. In Case 1, the true mode is trivial to find and it is the optimal threshold. For Case 4, the mode needs to be found numerically, using the expression for the density function. Once the mode is known, we define the candidate threshold set as sample quantiles at levels $0\%, \ldots, 95\%$ of the data which lies above the mode.

First, we consider the effect of the range $[u_1,u_k]$ of the candidate thresholds on threshold selection performance. Table~\ref{tab:modeC1C4} provides the RMSE, bias and variance of threshold estimation for samples from both of these cases, using the EQD, Wadsworth and Northrop methods each applied with candidate thresholds across the distribution, as given in Section~\ref{section: simstudyoverview} of the main paper, and only above the true mode. In terms of RMSE, the EQD method outperforms the other two methods regardless of how the candidate threshold grid is chosen. For the ``Above mode" grid, we have rather different performances for the Wadsworth and Northrop methods relative to the ``Original" candidate grid. In Case 1, these two methods perform very similarly, with RMSEs approximately 15 times larger than the EQD. In Case 4, the Wadsworth method fails completely with the ``Above mode" grid due to the grid being too fine relative to the sample size. For the Northrop method, the RMSE of threshold estimation is increased with the ``Above mode" grid and the differential between the Northrop and EQD becomes wider with a RMSE that is 1.15 times larger the EQD method. 

\begin{table}[h]
    \centering
    \scalebox{0.9}{
    \begin{tabular}{|c|c|c|c|c|c|c|c|c|c|}
    \hline
         & RMSE & Bias & Variance & RMSE & Bias & Variance & RMSE & Bias & Variance \\
         \hline
         & \multicolumn{9}{c|}{Case 1}\\
        \hline
        & \multicolumn{3}{c|}{EQD} & \multicolumn{3}{c|}{Wadsworth} & \multicolumn{3}{c|}{Northrop}\\
        \hline
        Original & 0.048 & 0.034 & 0.001 & 0.349 & 0.111 & 0.110 & 0.536 & 0.276 & 0.212 \\
        Above mode & 0.038 & 0.018 & 0.001 & 0.577 & 0.236 & 0.277 & 0.597 & 0.348 & 0.235\\
        \hline
        & \multicolumn{9}{c|}{Case 4}\\
        \hline
        Original & 0.526 & $-0.515$ & 0.012 & 0.627 & $-0.407$ & 0.230 & 0.543 & $-0.222$ & 0.246 \\
        Above mode & 0.514 & $-0.505$ & 0.009 & NA & NA & NA & 0.589 & $-0.127$ & 0.331 \\
        \hline
    \end{tabular}}
    \caption{RMSE, bias and variance of threshold choice for Case 1 and Case 4 samples for EQD, Wadsworth and Northrop methods. The methods are evaluated for two sets of candidate threshold sets: Original, using candidate threshold across the whole sample;  Above mode, using  candidate thresholds which only lie above the mode.}
    \label{tab:modeC1C4}
\end{table}

For the EQD method, using only candidate thresholds above the true mode results in a smaller RMSE (and bias and variance) for both Cases 1 and 4, presumably as we are now using a finer grid of candidate thresholds so candidate thresholds lie closer to the optimal value. In contrast, the performance of the Wadsworth and Northrop methods is worsened when we restrict the candidate thresholds to be above the mode, primarily due to an increase in variance.

We also consider the effect of the range of the candidate thresholds on quantile estimation for the Gaussian case. Previously, for the Gaussian case, we had reported results only using candidate threshold above the sample median of the data, i.e., essentially the mode. Here, we assess the effect of allowing candidate thresholds to range across the whole sample, i.e., the sample quantiles at levels $0\%, 5\%, \ldots, 95\%$ of the whole data sample.

The results, in terms of RMSE for three different quantiles, are presented in Table~\ref{tab: gauss_mode} for the two different candidate threshold grids. Firstly, we find that the Wadsworth method completely fails with this extended range of candidate thresholds, so this method is omitted from the table. For both the EQD and Northrop methods, we have almost identical values regardless of the set of candidate thresholds we use, with the RMSE always smallest for the EQD.

\begin{table}[h]
    \centering
    \begin{tabular}{|c|c|c|c|c|}
    \hline
    & \multicolumn{2}{c|}{EQD} & \multicolumn{2}{c|}{Northrop}\\
    \hline
     $j$ & Original & Whole data  & Original & Whole data \\
     \hline
     0 & 0.214 & 0.214 & 0.225 & 0.225 \\
     1 & 0.430 & 0.431 & 0.461 & 0.460 \\
     2 & 0.703 & 0.707 & 0.765 & 0.763 \\
     \hline
    \end{tabular}
    \caption{RMSEs of estimated $(1-p_{j,n})$-quantiles where $p_{j,n}=1/(10^jn)$, for $j=0,1,2$, for the Gaussian case when applying methods with candidate thresholds across the range of the data.}
    \label{tab: gauss_mode}
\end{table}

The results here indicate that the choice of candidate grid taken in the main paper provides no unfair advantage to the EQD method, in fact it performs even better if the choice of candidate threshold grid is tuned by exploiting knowledge of the value of the mode of the distribution.

\section{Further details on coverage probability results}\label{section: further_coverage}

This section provides more detailed results for the coverage probabilities for the 500 replicated samples of Case 4 and the Gaussian case outlined in Section~\ref{section: simstudy} of the main text. Here, we also show results for the 50\% confidence level and include an extended range of exceedance probabilities at which the coverage of the true quantiles is evaluated. 

Tables~\ref{tab: coverage_case4_full} and \ref{tab: coverage_norm_full} expanding on the results given in Tables~\ref{tab: coverage_case4} and \ref{tab: coverage_norm} of the main paper, providing coverage probabilities and average CI width ratios using Algorithms 1 and 2 for 500 samples derived from Case 4 and the Gaussian distribution. These results allow us to conclude that the additional threshold uncertainty, captured by Algorithm~2, leads to a significant improvement in the calibration of CIs and that it is vital to include this uncertainty in any inference for extreme quantiles. 

\textbf{Scenario 1: True GPD tail - Case 4}\\
Table~\ref{tab: coverage_case4_full} shows that for all confidence levels and exceedance probabilities, Algorithm 1, which includes  uncertainty only from the GPD parameter estimation, substantially under-estimates the uncertainty in quantile estimates. This leads to CIs which do not cover the true quantiles to the nominal levels of confidence. 
In contrast, the inclusion of the additional threshold uncertainty in Algorithm 2 leads to significant increases in the coverage of true quantiles with coverage probabilities at all quantile levels lying very close to the nominal confidence level, especially at the 95\% level. From a practical perspective, it is reassuring that this improvement in coverage is achieved with only 40-68\% average increases in the width of the CIs.

\begin{table}[ht]
    \centering
    \begin{tabular}{|c|c|c|c|c|c|c|c|c|c|c|} 
     \hline
       $p$ & $1/n$ & $1/3n$ & $1/5n$ & $1/10n$ & $1/25n$ & $1/50n$ & $1/100n$ & $1/200n$ & $1/500n$ \\
         \hline
     & \multicolumn{9}{c|}{50\% confidence}\\
       \hline
     \textit{Alg 1} & 0.398 & 0.390 & 0.384 & 0.368 & 0.368 & 0.358 & 0.358 & 0.350 & 0.342\\
     \textit{Alg 2} & 0.526 & 0.498 & 0.500 & 0.496 & 0.480 & 0.476 & 0.474 & 0.468 & 0.464  \\
     \textit{CI ratio} & 1.408 & 1.413 & 1.414 & 1.414 & 1.414 & 1.413 & 1.413 & 1.412 & 1.411\\
      \hline
       & \multicolumn{9}{c|}{80\% confidence}\\
       \hline
     \textit{Alg 1} & 0.646 & 0.642 & 0.630 & 0.618 & 0.616 & 0.608 & 0.606 & 0.600 & 0.594 \\
     \textit{Alg 2} & 0.798 & 0.778 & 0.770 & 0.772 & 0.760 & 0.760 & 0.758 & 0.762 & 0.758 \\
     \textit{CI ratio} & 1.430 & 1.440 & 1.445 & 1.452 & 1.461 & 1.468 & 1.475 & 1.483 & 1.495\\
      \hline
       & \multicolumn{9}{c|}{95\% confidence}\\
       \hline
     \textit{Alg 1} & 0.834 & 0.810 & 0.808 & 0.804 & 0.798 & 0.796 & 0.794 & 0.788 & 0.788 \\
     \textit{Alg 2} & 0.954 & 0.950 & 0.950 & 0.948 & 0.942 & 0.942 & 0.944 & 0.944 & 0.944  \\
     CI ratio & 1.484 & 1.511 & 1.525 & 1.546 & 1.574 & 1.597 & 1.621 & 1.646 & 1.682 \\
      \hline
    \end{tabular}
    \caption{Coverage probabilities for estimated $(1-p)$-quantiles using Algorithms 1 and 2  
    for Case 4, with sample size of 1000. Values are based on 500 replicated samples.}
    \label{tab: coverage_case4_full}
\end{table}

\FloatBarrier

\textbf{Scenario 2: Gaussian data}\\
Table~\ref{tab: coverage_norm_full}
shows that, for Gaussian variables across all quantiles, Algorithm 1 and 2 are less successful in the coverage of the true quantiles than in Case 4.  Specifically, there is a significant under-estimation of the estimated uncertainty necessary to provide coverage probabilities near to the nominal confidence level, and the actual coverage is
decreasing with the level of extrapolation required.
This is not too surprising as it is well-established that Gaussian variables exhibit quite slow convergence to an extreme value limit. However, what Table~\ref{tab: coverage_norm_full} shows is that
the additional threshold uncertainty in Algorithm~2 leads to a substantial improvement in actual coverage across all quantiles and for all nominal confidence levels. This improvement is achieved with the CI widths on average being extended by 45-73\%. In particular, for the $p=1/n, 1/3n, 1/5n$, which are typical levels of extrapolation from a sample in practical contexts, Algorithm~2 achieves a workable performance, with coverage reasonably close to the nominal level.
\begin{table}[ht]
    \centering
    \begin{tabular}{|c|c|c|c|c|c|c|c|c|c|c|} 
     \hline
       $p$ & $1/n$ & $1/3n$ & $1/5n$ & $1/10n$ & $1/25n$ & $1/50n$ & $1/100n$ & $1/200n$ & $1/500n$ \\
         \hline
     & \multicolumn{9}{c|}{50\% confidence}\\
       \hline
     \textit{Alg 1} &  0.358 & 0.300 & 0.292 & 0.278 & 0.240 & 0.218 & 0.204 & 0.190 & 0.168\\
     \textit{Alg 2} & 0.462 & 0.398 & 0.354 & 0.322 & 0.278 & 0.262 & 0.232 & 0.212 & 0.200  \\
     CI ratio & 1.461 & 1.463 & 1.462 & 1.465 & 1.465 & 1.466 & 1.465 & 1.466 & 1.465\\
      \hline
       & \multicolumn{9}{c|}{80\% confidence}\\
       \hline
     \textit{Alg 1} & 0.588 & 0.522 & 0.498 & 0.450 & 0.402 & 0.388 & 0.366 & 0.348 & 0.316 \\
     \textit{Alg 2} & 0.718 & 0.656 & 0.630 & 0.598 & 0.542 & 0.516 & 0.492 & 0.476 & 0.446 \\
     CI ratio & 1.457 & 1.468 & 1.473 & 1.480 & 1.493 & 1.501 & 1.509 & 1.517 & 1.526 \\
      \hline
       & \multicolumn{9}{c|}{95\% confidence}\\
       \hline
     \textit{Alg 1} & 0.750 & 0.674 & 0.650 & 0.618 & 0.580 & 0.550 & 0.510 & 0.490 & 0.466\\
     \textit{Alg 2} & 0.866 & 0.838 & 0.828 & 0.814 & 0.794 & 0.772 & 0.756 & 0.742 & 0.722 \\
     CI ratio & 1.495 & 1.531 & 1.549 & 1.576 & 1.611 & 1.638 & 1.665 & 1.692 & 1.729 \\
      \hline
      
    \end{tabular}
    \caption{Coverage probabilities for estimated $(1-p)$-quantiles using Algorithms 1 and 2  
    for a Gaussian distribution, with sample size of 2000. Values are based on 500 replicated samples.}  
    \label{tab: coverage_norm_full}
\end{table}



\end{document}